\documentclass[9pt,twocolumn,twoside]{pnas-new}
\setboolean{displaywatermark}{false}

\templatetype{pnasresearcharticle}
\usepackage{upgreek}
\DefineJournalPartialAbbreviation{Atmospher}{Atmos}

\title{The rotational and divergent components of atmospheric circulation on tidally locked planets}

\author[a,1]{Mark Hammond}
\author[b,1,2]{Neil T. Lewis} 

\affil[a]{Department of the Geophysical Sciences, University of Chicago, USA}
\affil[b]{Atmospheric, Oceanic and Planetary Physics, University of Oxford, UK}

\leadauthor{Hammond and Lewis} 

\significancestatement{Tidally locked exoplanets have a permanent day-side and night-side. Understanding the atmospheric circulation on these planets is crucial for interpreting telescope observations, and assessing their habitability.  We show that the main components of the circulation -- a jet going around the planet, stationary atmospheric waves, and direct flow from the day-side to the night-side -- can be separated using a simple mathematical decomposition. This decomposition will significantly aid future study of tidally locked atmospheres. As an illustration, we use it to analyse heat transport by different components of the circulation. In contrast with previous work, we show that the direct day-night component can dominate heat transport from the day-side to the night-side, even when the jet is strong.}

\authorcontributions{M.H. and N.T.L. designed the study, conducted analysis, and wrote the manuscript.}
\authordeclaration{The authors declare no competing interests.}
\equalauthors{\textsuperscript{1}M.H. and N.T.L. contributed equally to this work.}
\correspondingauthor{\textsuperscript{2}To whom correspondence should be addressed. E-mail: neil.lewis@physics.ox.ac.uk}

\keywords{Exoplanets $|$ Atmospheric circulation $|$ Helmholtz decomposition $|$}

\begin{abstract}
Tidally locked exoplanets likely host global atmospheric circulations with a superrotating equatorial jet, planetary-scale stationary waves and thermally-driven overturning circulation. In this work, we show that each of these features can be separated from the total circulation by using a Helmholtz decomposition, which splits the circulation into rotational (divergence free) and divergent (vorticity free) components. This technique is applied to the simulated circulation of a terrestrial planet and a gaseous hot Jupiter. For both planets, the rotational component comprises the equatorial jet and stationary waves, and the divergent component contains the overturning circulation. Separating out each component allows us to evaluate their spatial structure and relative contribution to the total flow. In contrast with previous work, we show that divergent velocities are not negligible when compared with rotational velocities, and that divergent, overturning circulation takes the form of a single, roughly isotropic cell that ascends on the day-side and descends on the night-side. These conclusions are drawn for both the terrestrial case and the hot Jupiter. To illustrate the utility of the Helmholtz decomposition for studying atmospheric processes, we compute the contribution of each of the circulation components to heat transport from day- to night-side. Surprisingly, we find that the divergent circulation dominates day-night heat transport in the terrestrial case and accounts for around half of the heat transport for the hot Jupiter. The relative contributions of the rotational and divergent components to day-night heat transport are likely sensitive to multiple planetary parameters and atmospheric processes, and merit further study. 
\end{abstract}

\dates{Submitted 30 October 2020; Revised 16 February 2021; Accepted 19 February 2021}
\doi{\url{www.pnas.org/cgi/doi/10.1073/pnas.XXXXXXXXXX}}

\begin{document}

\maketitle
\thispagestyle{firststyle}
\ifthenelse{\boolean{shortarticle}}{\ifthenelse{\boolean{singlecolumn}}{\abscontentformatted}{\abscontent}}{}

\dropcap{E}xoplanets, which are planets orbiting stars other than the Sun, have revealed a variety of novel forms of atmospheric circulation. The most notable of these is the circulation of tidally locked planets, which are close enough to their host star that tidal stresses between planet and star cause the planet's orbital and rotational periods to synchronise \cite{dole1964locking,guillot1996locking}. These planets always present the same side to their star, yielding a permanent day-side and night-side.

Understanding the global circulation of tidally locked planets is vital to interpreting observations of their atmospheres, and studying their atmospheric stability and habitability. The circulation transports heat, chemical species, and clouds around the planet. This affects the observable `phase curve', which is the light emitted or reflected by the planet as it rotates \cite{burrows2014review,crossfield2015observations,parmentier2017handbook}. 
Accurate retrievals of chemical composition and cloud structure depend on understanding the temperature structure of the atmosphere, which is determined by the circulation \cite{feng2016emission,feng2020retrieval2d,irwin20202,taylor2020jwstspec}. Vertical motion in the atmosphere affects the transport and distribution of chemical species and clouds \cite{zhang2018globaltl,komacek2019vertical}, which will also have observable effects. In addition, the circulation may be crucial to supporting a habitable atmosphere on a terrestrial (rocky) planet, with sufficient heat transport needed to prevent volatile species from condensing on the cold night-side and leading to atmospheric collapse \cite{joshi1997tidally,heng2012collapse,wordsworth2015collapse,turbet2018modeling}.

Previous work has shown that the circulation on tidally locked planets is driven by the strong heating/cooling gradient between their day-sides and night-sides  \citep{showman2013review,heng2015review,pierrehumbert2018review,zhang2020atmospheric}. The day-night forcing generates overturning circulation, which features air rising on the day-side and sinking on the night-side \citep{showman2013review}. This vertical motion then leads to the generation of stationary waves \citep{sardeshmukh1988generation,showman2010superrotation}, which in turn can accelerate a prograde (superrotating) equatorial jet \citep{showman2010superrotation,showman2011superrotation,tsai2014three,hammond2018wavemean}. However, the relative contribution of each of these components to the total circulation is poorly understood, as no study has shown how they can be isolated from one another.

In this study we address this issue by showing how the overturning circulation, stationary waves, and superrotating jet can be separated out from the total circulation using a Helmholtz decomposition, which uniquely divides the total circulation $\boldsymbol{u}=(u,v)$ into divergent (`vorticity free') and rotational (`divergence free') components \cite{dutton1976ceaseless} \begin{linenomath*}
\begin{align}
    \boldsymbol{u}&=\boldsymbol{u}_{d}+\boldsymbol{u}_{r} \label{eq:decomp} \\
    &= -\nabla\chi+\mathbf{k}\times\nabla\psi.
\end{align}
\end{linenomath*}
Above, $\chi$ is the velocity potential function, and $\psi$ is a streamfunction. $\chi$ and $\psi$ are obtained from 
\begin{linenomath*}
\begin{align}
    \nabla^{2}\chi&=\delta \label{eq:chi}\\ 
    \nabla^{2}\psi&=\zeta, \label{eq:psi}
\end{align}
\end{linenomath*}
where $\delta$ is the divergence and $\zeta$ is the vorticity. We apply the Helmholtz decomposition to the horizontal velocity fields output from two well-studied General Circulation Model (GCM) simulations of tidally locked atmospheres; one representing a terrestrial planet \cite{hammond2020equatorial}, and the other a giant gaseous planet (a `hot Jupiter') \cite{deitrick2020thor}. Details of the numerical procedure used to invert Eq. \ref{eq:chi} and Eq. \ref{eq:psi} is provided in the Materials and Methods section. The Helmholtz decomposition has been used extensively to study the Earth's atmospheric circulation (see, e.g., \citealp{dutton1976ceaseless}). We apply it for the first time to the atmospheric circulation of tidally locked planets.

\begin{figure*}[!t]
\centering\includegraphics[width=0.45\linewidth]{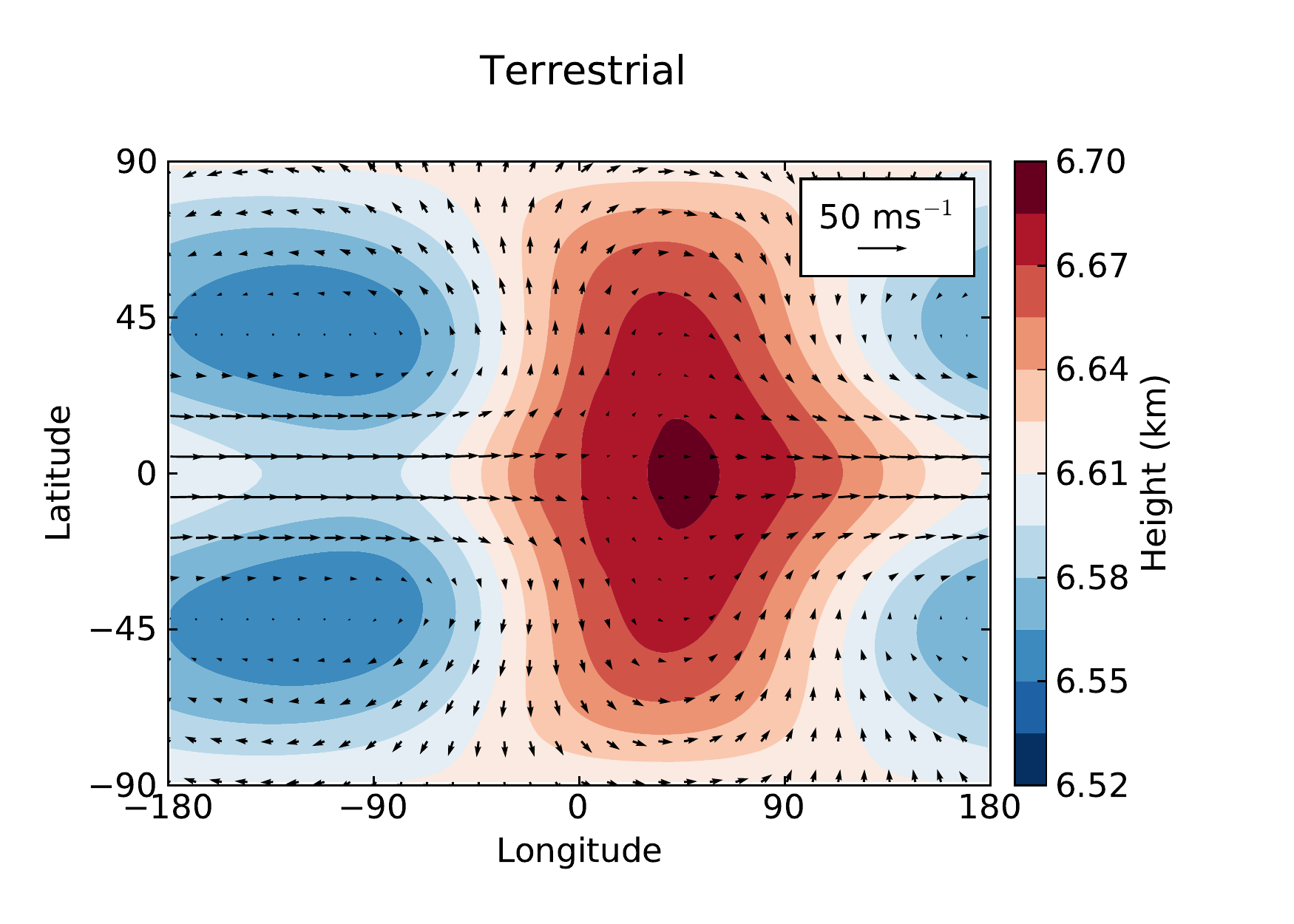}
\centering\includegraphics[width=0.45\linewidth]{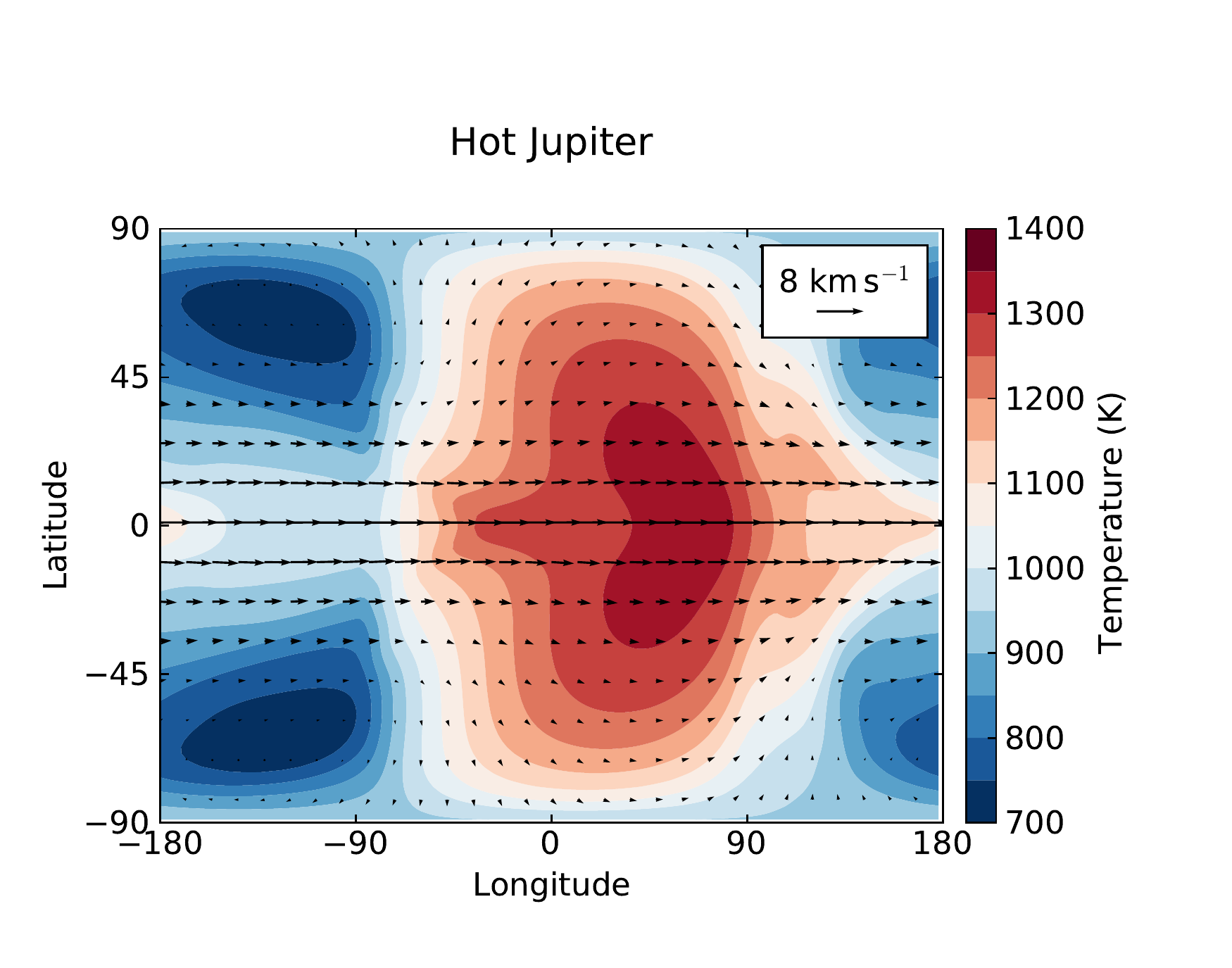} \\
\vspace*{-.2in}
\includegraphics[width=0.45\linewidth]{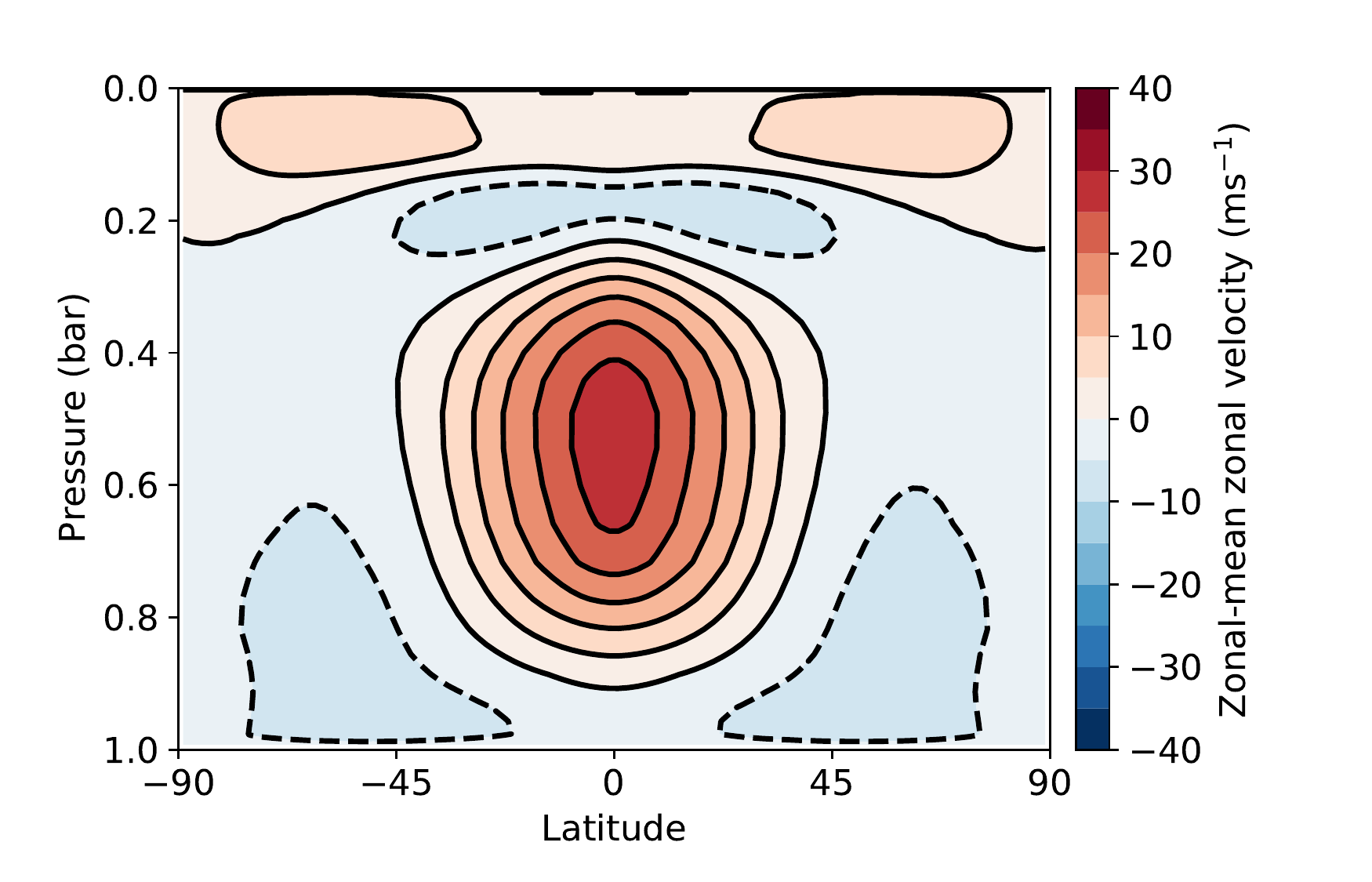} 
\includegraphics[width=0.45\linewidth]{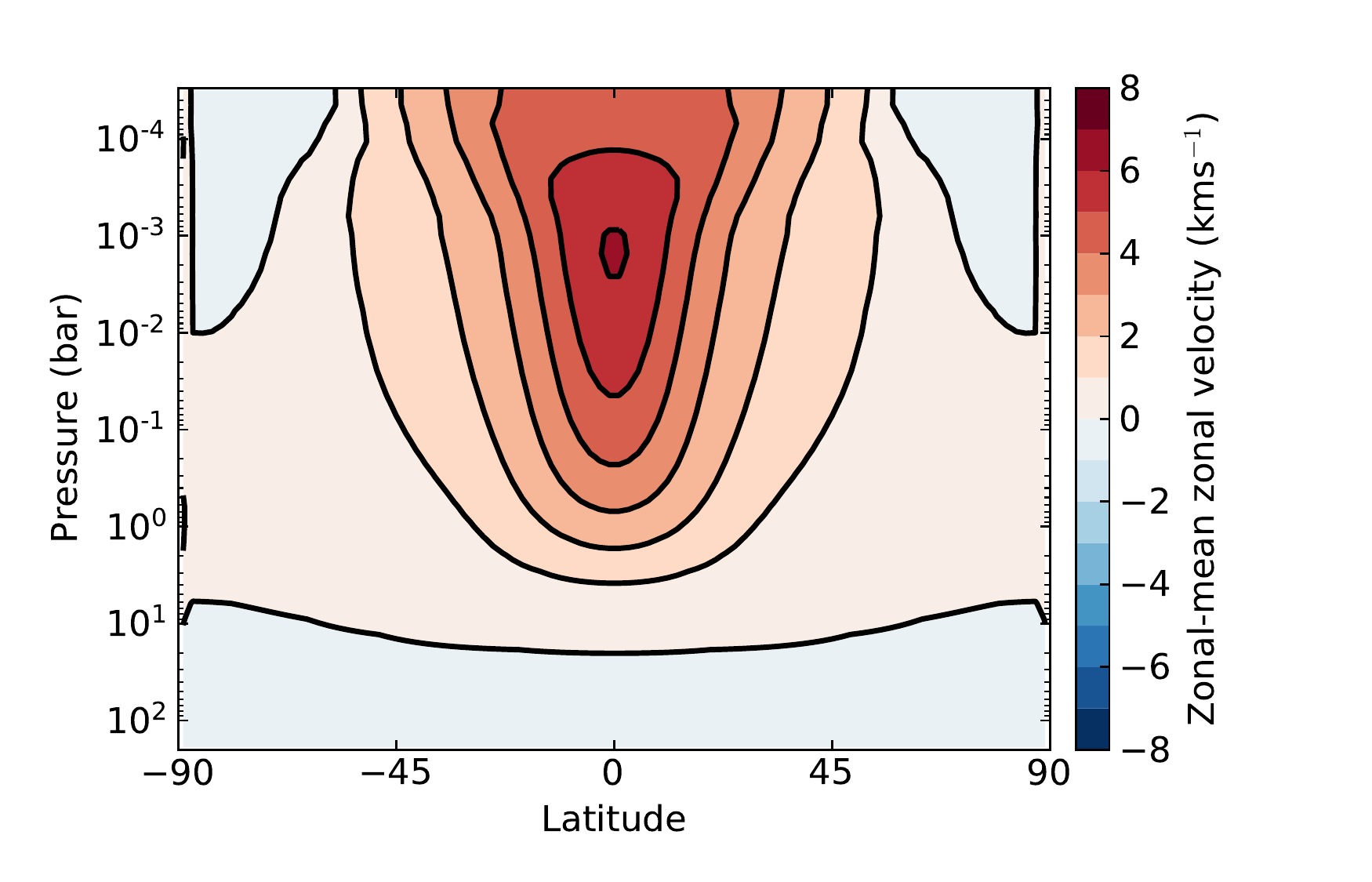}
\caption{The global circulation of the idealised tidally locked planets simulated in Exo-FMS and THOR. Left column: Terrestrial simulation, showing the height and velocity fields at 0.4 bar, and the zonal-mean zonal velocity. Right column: Hot Jupiter simulation, showing the temperature and velocity fields at 0.02 bar, and the zonal-mean zonal velocity. Both simulations have the eastward equatorial jet and eastward hot-spot shift typical to tidally locked planets. For both simulations, the substellar point is located at ($0^{\circ},0^{\circ}$). We show the height field for the terrestrial case and the temperature field for the gaseous case for consistency with the original publications where their overall circulation was analysed \cite{hammond2020equatorial,deitrick2020thor}. In both atmospheres, the height and temperature fields have the same qualitative structure as the thickness of a hydrostatic layer between two pressure levels is proportional to its temperature \cite{showman2011superrotation}. This relationship is expressed by the hypsometric equation (see, e.g., ref. \citealp{wallace2006atmospheric}).}\label{fig:terr-circulation}
\end{figure*}

Fig. \ref{fig:terr-circulation} shows the global circulation of the terrestrial planet and hot Jupiter simulations that we analyse in this study. The terrestrial simulation was run using the GCM Exo-FMS \cite{hammond2018wavemean,hammond2020equatorial}, using parameters appropriate for typical terrestrial tidally locked planets, such as those in the Trappist-1 system \cite{gillon2017seven}. The hot Jupiter simulation was run using the GCM THOR \cite{mendoncca2016thor,deitrick2020thor}, configured with parameters appropriate for the planet HD 189733b \cite{bouchy2005hd189773b}. In both simulations, the substellar point is located at $0^{\circ}$ longitude, $0^{\circ}$ latitude. Model details and parameters are described in Materials and Methods for both cases. The data for the THOR simulation was provided to us by the developers of THOR. 

The left-hand column of Fig. \ref{fig:terr-circulation} shows the circulation of the terrestrial planet simulation. The top panel shows the height and velocity field at $0.4\,\text{bar}$, and the bottom panel shows the zonal-mean zonal velocity. These fields show the key features of its circulation: a `hot-spot' shifted eastwards of the substellar point, stationary planetary-scale waves, and an eastward equatorial jet produced by these waves \citep{showman2011superrotation,tsai2014three,hammond2018wavemean}. The right-hand column of Fig. \ref{fig:terr-circulation} shows the hot Jupiter. Temperature and velocity fields at $0.02\,\text{bar}$ are shown in the top panel, and the zonal-mean zonal velocity is shown in the bottom panel. This atmosphere has the same key features as the terrestrial case, despite its much higher temperature, larger size, and faster rotation rate. In this study we will decompose the velocity fields of our two simulations into two physically distinct circulations and show how they relate to these key features.

\section*{Results}

\subsection*{Terrestrial Planet}

\begin{figure*}
\centering\includegraphics[width=0.45\linewidth]{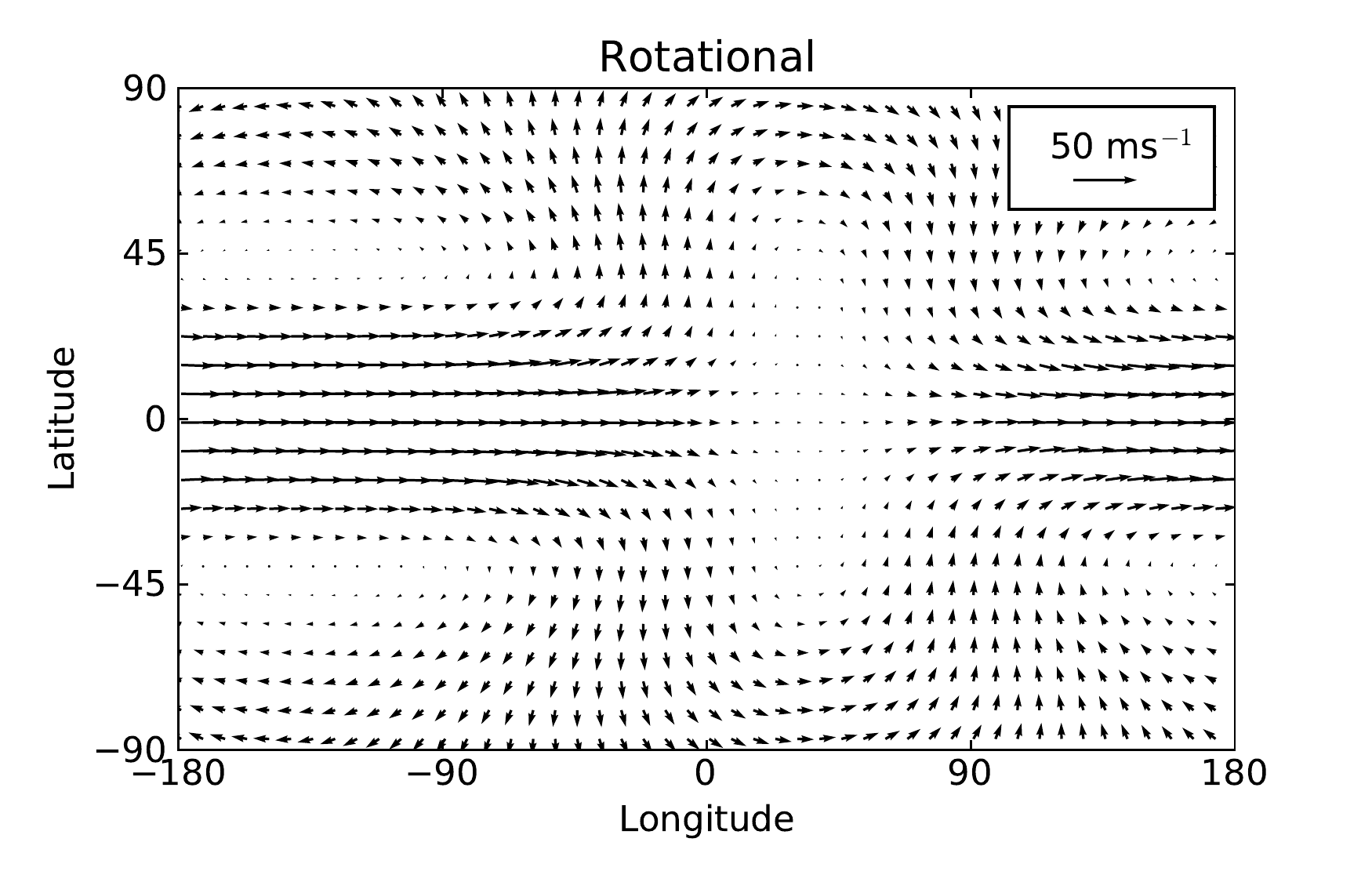}
\hspace*{.35in}\includegraphics[width=0.45\linewidth]{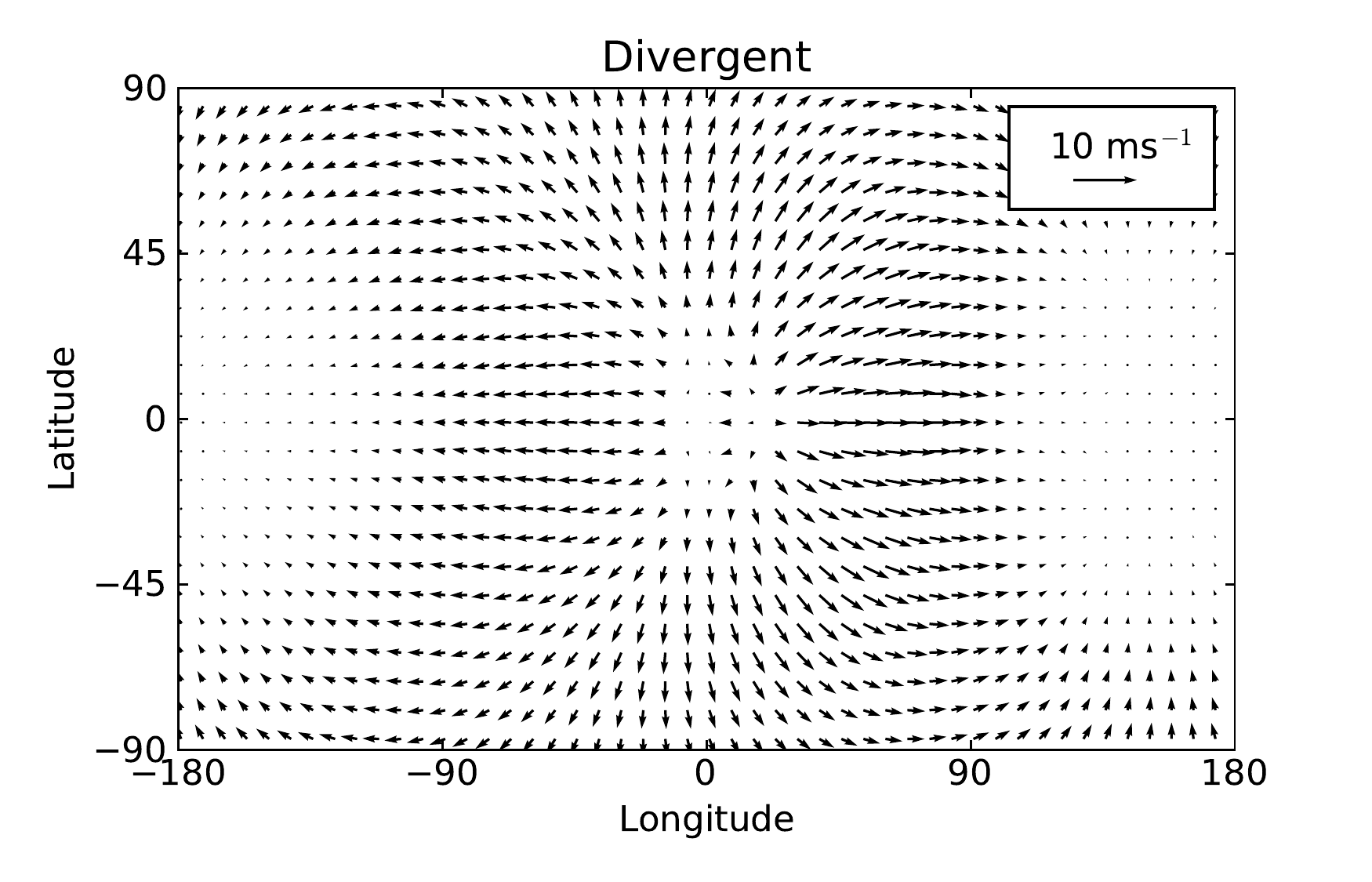}
\caption{Helmholtz decomposition of horizontal velocity $\boldsymbol{u}$ for the terrestrial simulation at 0.4 bar. Left panel: Rotational component of $\boldsymbol{u}$, $\boldsymbol{u}_{\text{r}}$. Fig. \ref{fig:rot} shows that this component is composed of the zonal-mean eastward equatorial jet and a stationary wave with zonal wavenumber 1. Right panel: Divergent component of $\boldsymbol{u}$, $\boldsymbol{u}_{\text{d}}$. At the pressure level shown here, the divergent component is dominated by an isotropic flow away from the substellar point ($0^{\circ},0^{\circ}$), which we will show to be associated with a single overturning cell.}\label{fig:decomp}
\end{figure*}

Fig. \ref{fig:decomp} shows $\boldsymbol{u}_{\text{r}}$ and $\boldsymbol{u}_{\text{d}}$ at $0.4\,\text{bar}$, calculated from Eqs. \ref{eq:decomp}--\ref{eq:psi} for the terrestrial circulation shown in Fig. \ref{fig:terr-circulation}. The rotational circulation is characterised primarily by zonal flow around the planet's axis of rotation, superimposed over a stationary wave pattern, and the divergent circulation is characterised by roughly isotropic flow away from the substellar point. 

\begin{figure}[!t]
    \centering
    \includegraphics[width=0.90\linewidth]{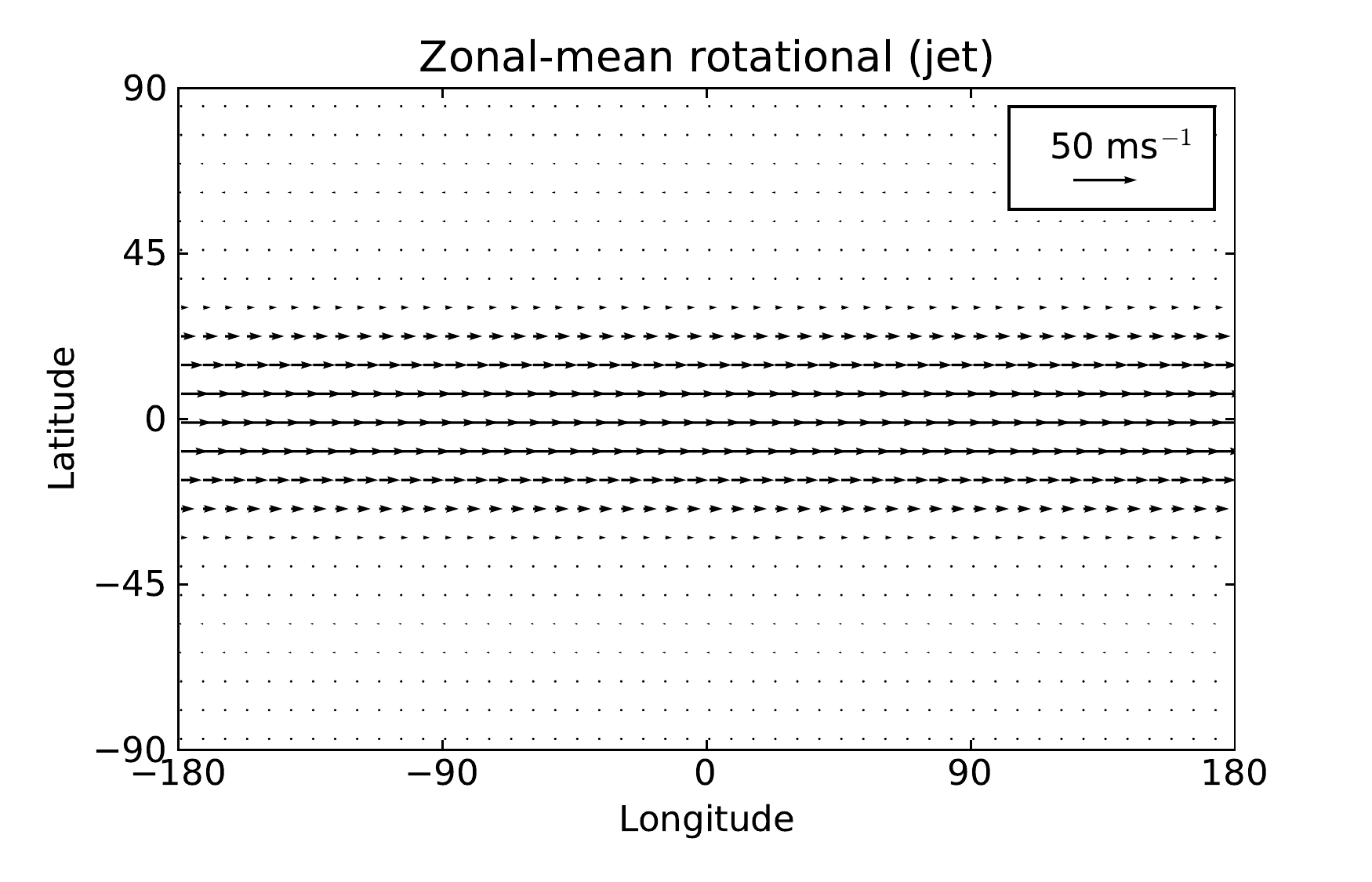} \\ \includegraphics[width=0.90\linewidth]{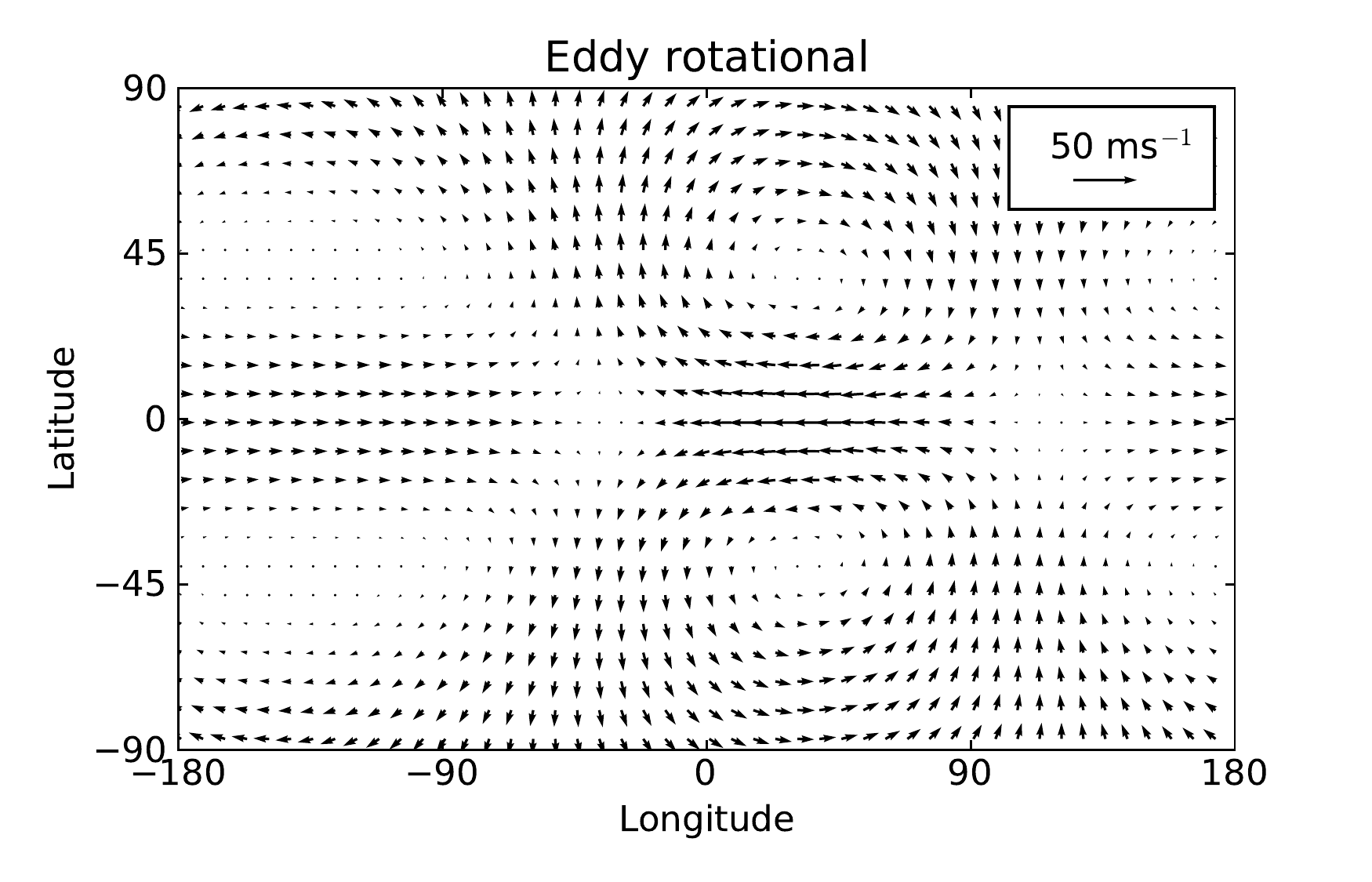}
    \caption{The two physical components of the rotational circulation in Fig. \ref{fig:decomp}. Top panel: Zonal-mean part of $\boldsymbol{u}_{\text{r}}$, $\overline{\boldsymbol{u}}_{\text{r}}$, which is the eastward equatorial jet. Bottom panel: Eddy (total minus zonal mean) part of $\boldsymbol{u}_{\text{r}}$, $\boldsymbol{u}_{\text{r}}^{\prime}$, which is dominated by a wave with zonal wavenumber 1.}
    \label{fig:rot}
\end{figure}

The rotational circulation has two phenomenological components, which are manifest in the zonal-mean of $\boldsymbol{u}_{\text{r}}$ and the eddy component of $\boldsymbol{u}_{\text{r}}$ (shown in Fig. \ref{fig:rot}). The eddy component corresponds to planetary waves, primarily stationary Rossby waves with zonal wavenumber 1 \cite{showman2011superrotation,tsai2014three,hammond2018wavemean}, which are driven by the divergent circulation \cite{sardeshmukh1988generation}. The zonal-mean component of  $\boldsymbol{u}_{\text{r}}$ corresponds to the zonal-mean superrotating jet, which is produced by momentum transport towards the equator associated with the stationary waves \cite{showman2011superrotation,hammond2020equatorial}. There is no zonal-mean meridional rotational velocity because this velocity is $\mathbf{u}_{\text{r}} = \mathbf{k} \times \nabla\psi$, so the meridional part is $\partial\psi / \partial x$ which vanishes when integrated over longitude (assuming there is no surface topography to produce a discontinuity in $\psi$).

Existing theory of the global circulation of tidally locked planets has been built largely on the results of `shallow-water models', which represent the first baroclinic vertical mode of the atmosphere as a single fluid layer. Showman and Polvani \cite{showman2011superrotation} showed how forcing a shallow-water system with a spatially periodic heating (representing the day-night heating contrast) produces stationary waves. These waves produce an eastward acceleration at the equator in the shallow-water model, suggesting that the jet in fully 3D atmospheres forms in the same way. Tsai et al. \cite{tsai2014three} and Hammond and Pierrehumbert \cite{hammond2018wavemean} included the effect of the eastward jet on the stationary waves in shallow-water models, which explained how the aforementioned hot-spot shift forms on these planets. Each of these shallow-water models assume that the atmosphere is dominated by a stationary wave response with zonal wavenumber 1 and a zonal-mean zonal flow, and that these two features have separable effects. They also do not include any aspects of the overturning circulation. Fig. \ref{fig:rot} is therefore an important confirmation of the assumptions made in these shallow-water models as it shows that their stationary wave response corresponds to a specific part of the rotational circulation, which is indeed dominated by a component with zonal wavenumber 1. It also suggests that the overturning circulation is a physically distinct process from both the jet and stationary waves, so that it is reasonable to model the latter components in isolation in a shallow-water system.

\begin{figure}
    \centering
    \includegraphics[width=.90\linewidth]{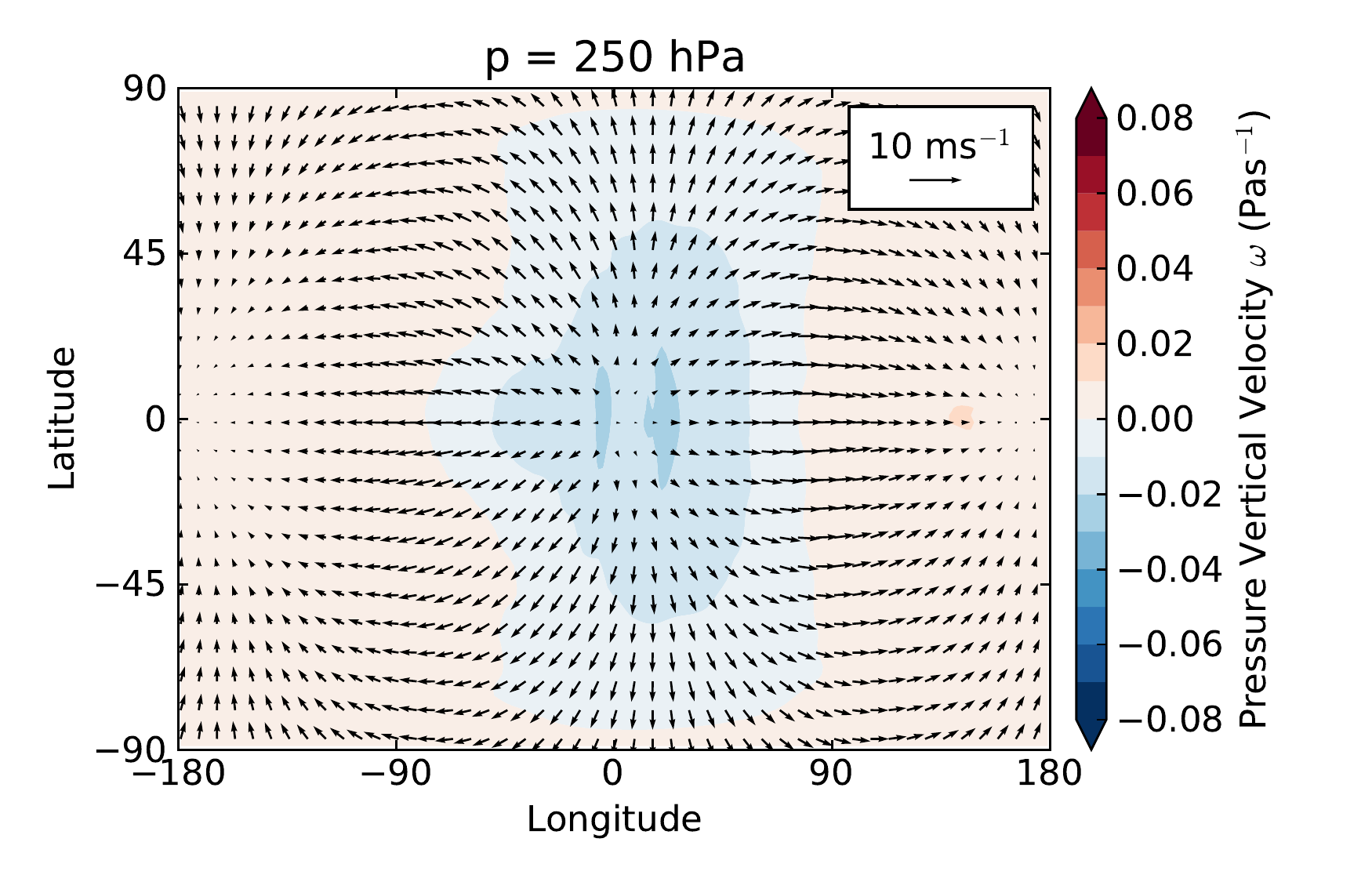} \\ 
    \includegraphics[width=.90\linewidth]{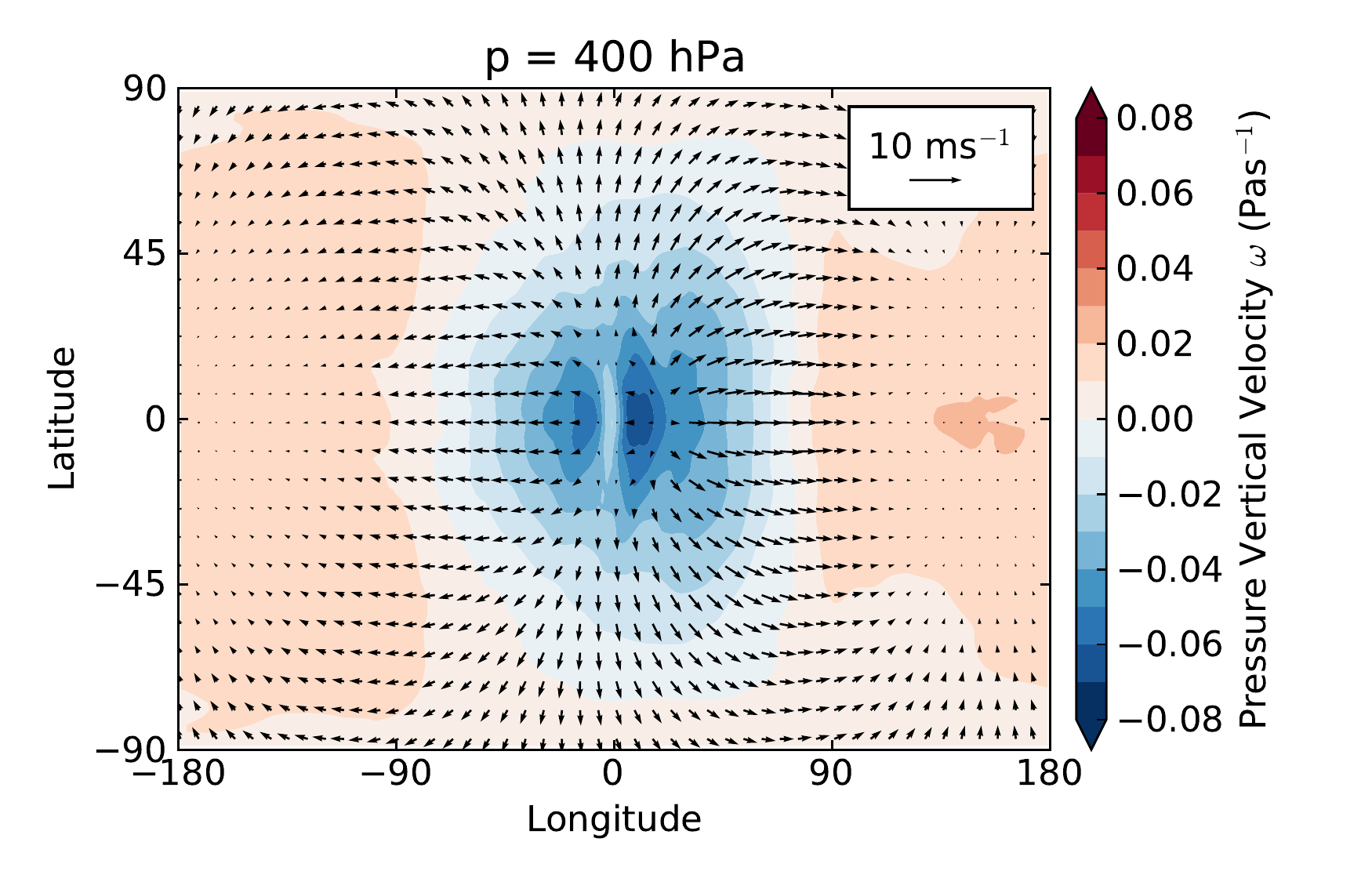} \\ 
    \includegraphics[width=.90\linewidth]{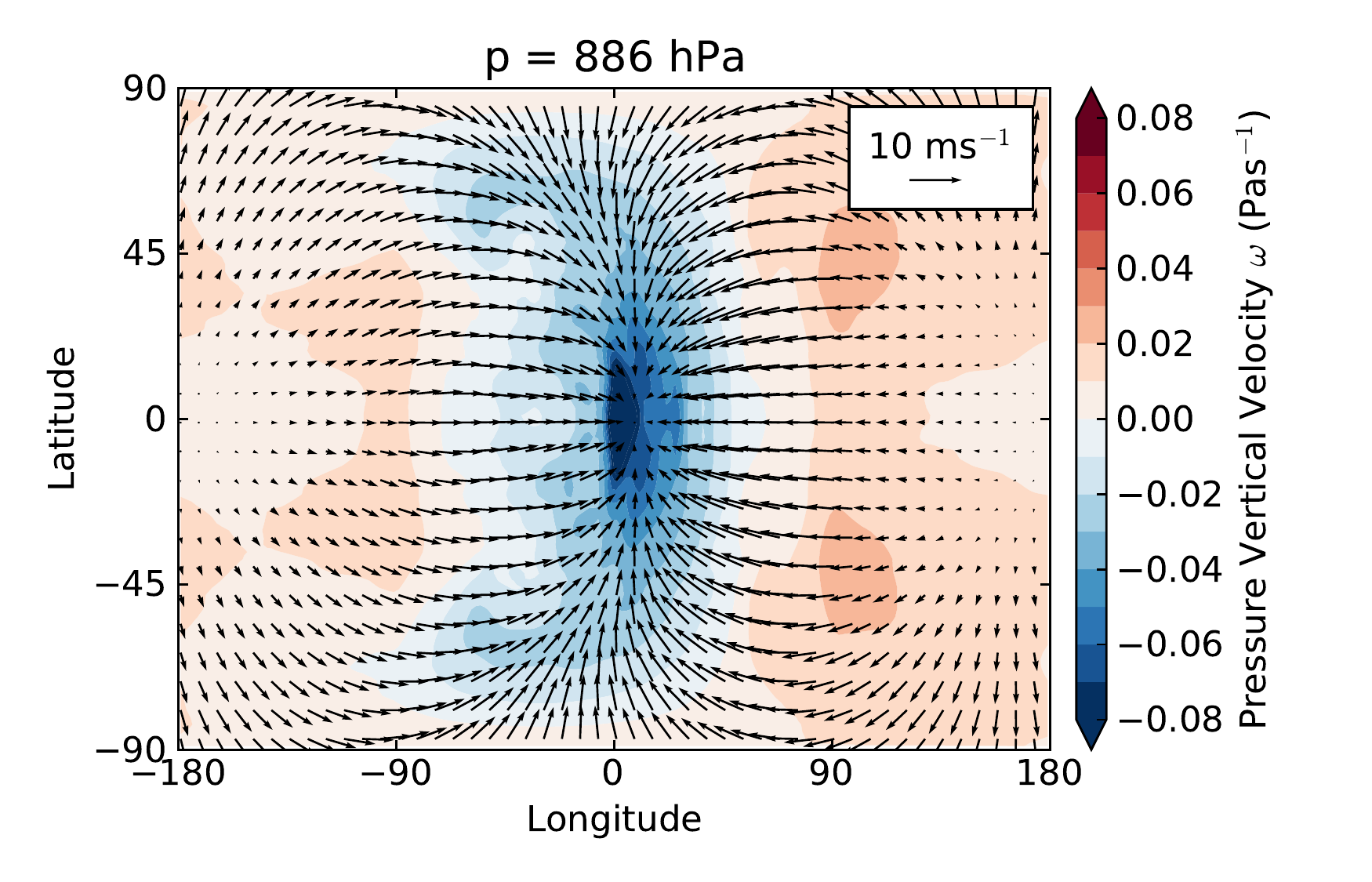} \\ 
    \includegraphics[width=.90\linewidth]{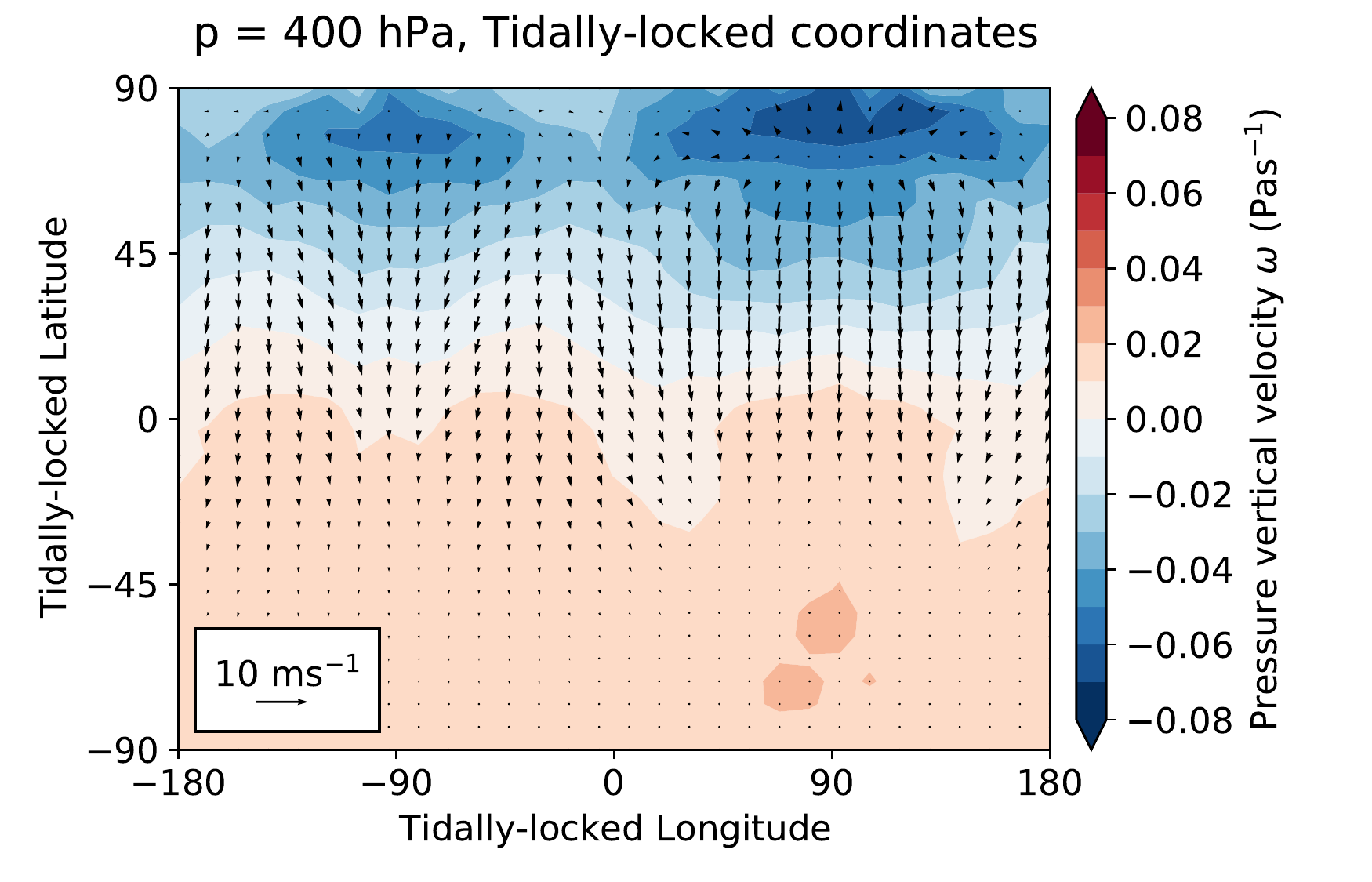}
    \caption{The divergent circulation of the terrestrial case. Each panel shows vertical velocity (shaded) and horizontal velocity (quivers). First to third panels: Latitude-longitude slices at $250\,\text{hPa}$ (first panel), $400\,\text{hPa}$ (second panel), and $886\,\text{hPa}$ (third panel), showing the day-night overturning cell. Fourth panel: Slice at $400\,\text{hPa}$ shown in tidally locked coordinates (see text), showing how the divergent flow is primarily along lines of constant tidally locked longitude in this coordinate system. }
    \label{fig:div}
\end{figure}

Divergent horizontal motion occurs due to convergent vertical motion (ignoring non-hydrostatic effects), and, phenomenologically speaking, the divergent circulation in our tidally locked simulation corresponds to an overturning circulation driven by heating at the substellar point. The divergent horizontal wind is shown in Fig. \ref{fig:div} at three different pressure levels: $250\,\text{hPa}$, $400\,\text{hPa}$, and $886\,\text{hPa}$, with the pressure vertical velocity $\omega$ underlaid as color contours (the quivers at $400\,\text{hPa}$ are the same as those in Fig. \ref{fig:decomp}). In the free atmosphere ($250\,\text{hPa}$ and $400\,\text{hPa}$), divergent horizontal motion is primarily a flow away from the substellar point, where radiative heating is strongest and air rises (negative $\omega$) and diverges. This flow extends roughly isotropically away from the substellar point in all directions, before starting to subside on the night-side (positive $\omega$). In the boundary layer ($886\,\text{hPa}$), air returns from regions of subsidence on the night-side towards the substellar point. The convergent return flow in the boundary layer is roughly isotropic, like the divergent flow aloft. There is some anisotropy in the boundary layer ascent (i.e., $\omega$) near the substellar point, which takes the form of a wide, horseshoe-shaped feature. This feature appears similar in structure to a weather front in the Earth's atmosphere \cite{hoskins1982fronts}, and forms at the boundary where cold air, advected from the night-side by the rotational circulation (which is predominantly westward in the boundary layer), meets the warm air in the vicinity of the substellar point.

At $400\,\text{hPa}$, it is notable that there are two main regions of convergence just beyond each terminator, rather than a single region at the antistellar point, suggesting that the overturning circulation is not sufficiently strong to reach all the way to the antistellar point in the mid-troposphere for the specific terrestrial planet considered here. Higher up in the atmosphere at $250\,\text{hPa}$, divergent flow extends further and convergence primarily occurs near the antistellar point. The divergent circulation is slightly stronger going east from the substellar point, compared to its westward part, likely due to the presence of the zonal-mean eastward equatorial jet; notably, this effect is more pronounced at $400\,\text{hPa}$ where the jet is stronger.

Past interpretation of the overturning circulation on tidally locked planets has described the circulation as a superposition of various \emph{different} overturning circulations. Ascent on the day-side paired with meridional horizontal motion and descent near the poles has been described as a sort of Hadley circulation \cite{merlis2010atmospheric,edson2011circulations,heng2011atmospheric,carone2014connecting,haqqmisra2018regimes}, and overturning circulation in the zonal direction has been described as a Walker circulation \cite{carone2014connecting,haqqmisra2018regimes}, by analogy with their namesakes in the Earth's tropical atmosphere. On the night-side, meridional motion has been described as an `anti-Hadley circulation' \cite{heng2011atmospheric,haqqmisra2018regimes}. Our analysis of the divergent circulation suggests that this interpretation of overturning in a tidally locked atmosphere is overly complicated, and misleading; the overturning circulation is not three different circulations, but instead a single thermally-direct and (roughly) isotropic circulation from the day-side to the night-side.

Given the roughly isotropic nature of the overturning circulation about the substellar point, it is instructive to visualise it within a coordinate system that exploits symmetry about the axis connecting the substellar and antistellar points. We therefore adopt the `tidally locked coordinates' of Koll and Abbot (\citealp{koll2015phasecurves}; see Materials and Methods), where tidally locked latitude $\vartheta^{\prime}$ is measured as the angle from the terminator, and tidally locked longitude $\lambda^{\prime}$ is the angle about the axis connecting the substellar and antistellar points. Lines at constant tidally locked longitude (i.e., tidally locked meridians) are arcs that connect the substellar and antistellar points. Tidally locked velocities $u^{\prime}$ and $v^{\prime}$ are defined in the $\lambda^{\prime}$ and $\vartheta^{\prime}$ directions respectively. The point $(\vartheta^{\prime},\lambda^{\prime})=(0,0)$ is chosen to coincide with the north pole. We note that the Helmholtz decomposition (Eqs. \ref{eq:decomp}--\ref{eq:psi}) of the horizontal velocities is invariant under the coordinate system rotation from latitude-longitude to tidally locked coordinates. This follows from the invariance of the $\nabla^{2}$ operator under rotation, and the fact that as $\zeta$ and $\delta$ are scalar fields, they are also invariant under rotation by definition.

The bottom panel of Fig. \ref{fig:div} shows the divergent velocity $\boldsymbol{u}_{\text{d}}$ in the tidally locked coordinate system at $0.4\,\text{bar}$. Comparison with the second panel in Fig. \ref{fig:div}, which shows $\boldsymbol{u}_d$ in regular latitude-longitude coordinates, illustrates how moving to tidally locked coordinates transforms the data. The substellar point, which was a point on the equator in the latitude-longitude coordinate, is now a `pole' in the new coordinate system. Consequently, all substellar ascending vertical motion is now at $\vartheta^{\prime}=90^{\circ}$. The `equator' in the tidally locked coordinate system is the terminator. `North' of the terminator (i.e., on the day-side), there is ascending vertical motion, and `south' of the terminator (i.e., on the night-side), there is descending vertical motion. The dominance of substellar to antistellar flow is clear in this coordinate system, taking the form of `southward' flow along lines of constant $\lambda^{\prime}$. The isotropic nature of the divergent circulation appears in the form of approximate zonal symmetry in  $\lambda^{\prime}$.

By analogy with the continuity equation in regular $\lambda$--$\vartheta$ coordinates, the continuity equation in tidally locked coordinates is \begin{linenomath*}\begin{equation}
    \frac{1}{a\cos\vartheta^{\prime}}\frac{\partial u^{\prime}}{\partial\lambda^{\prime}}+\frac{1}{a\cos\vartheta^{\prime}}\frac{\partial }{\partial\vartheta^{\prime}}\left(v^{\prime}\cos\vartheta^{\prime}\right) + \frac{\partial\omega}{\partial p} = 0. \label{eq:cont_TL}
\end{equation}\end{linenomath*}
Symmetry in the $\lambda^{\prime}$ direction motivates integrating \eqref{eq:cont_TL} over $\lambda^{\prime}$, which enables the definition of a tidally locked meridional mass streamfunction \begin{linenomath*}\begin{equation}
    \Psi^{\prime} = \frac{2\pi a\cos\vartheta^{\prime}}{g}\int^{p}_{0}\left[v^{\prime}\right]_{\lambda^{\prime}}\text{d}p, \label{eq:sf}
\end{equation}\end{linenomath*}
where $[\cdot]_{\lambda^{\prime}}$ indicates an average over tidally locked longitude. $\Psi^{\prime}$ is the total northward (in the $\vartheta^{\prime}$ direction, i.e. towards the antistellar point) mass flux above a particular pressure $p$, and mass flux in the $\vartheta^{\prime}$--$p$ plane follows lines of constant $\Psi^{\prime}$.

\begin{figure*}[!t]
    \centering
    \includegraphics[width=.45\linewidth]{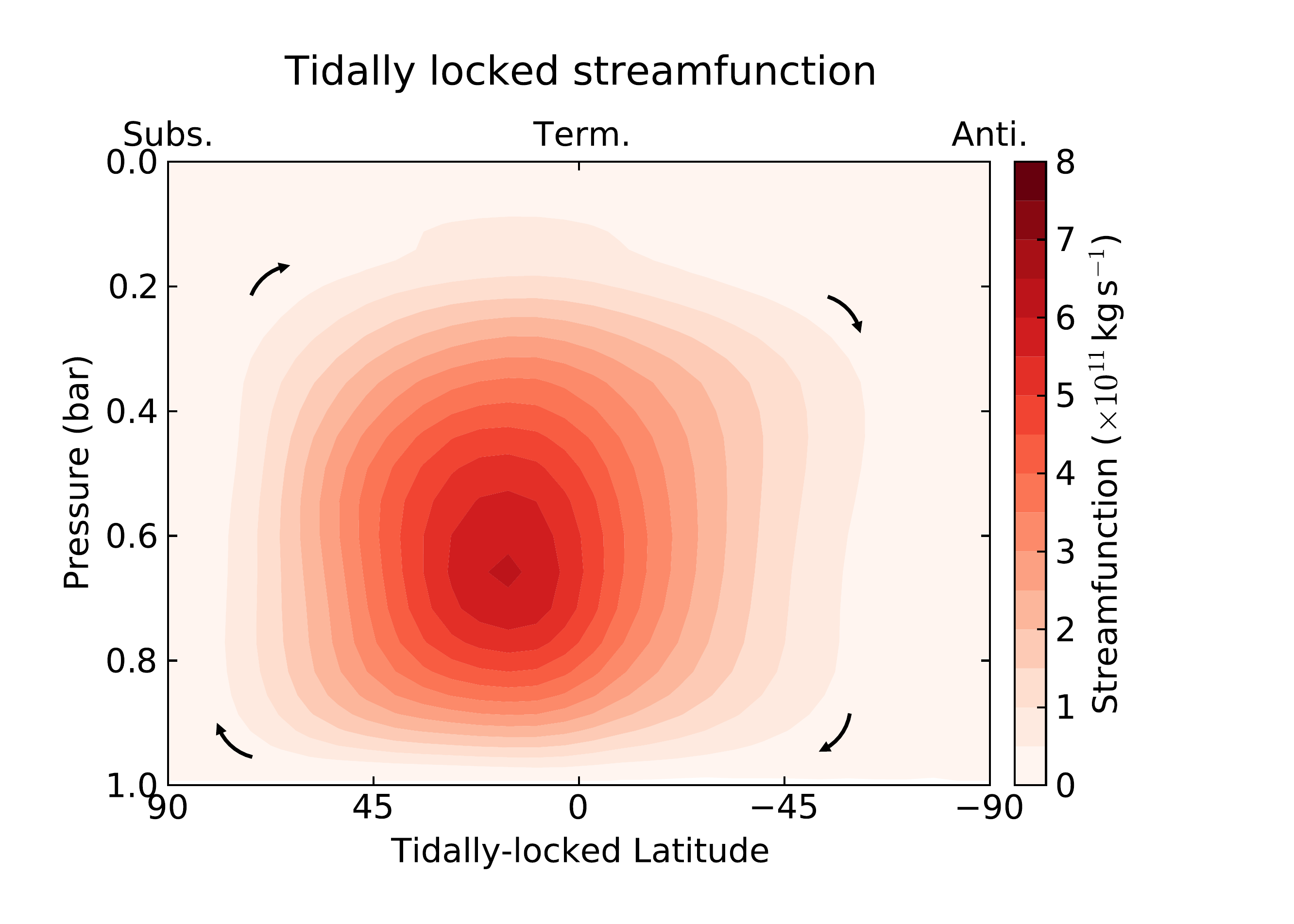} 
    \includegraphics[width=.45\linewidth]{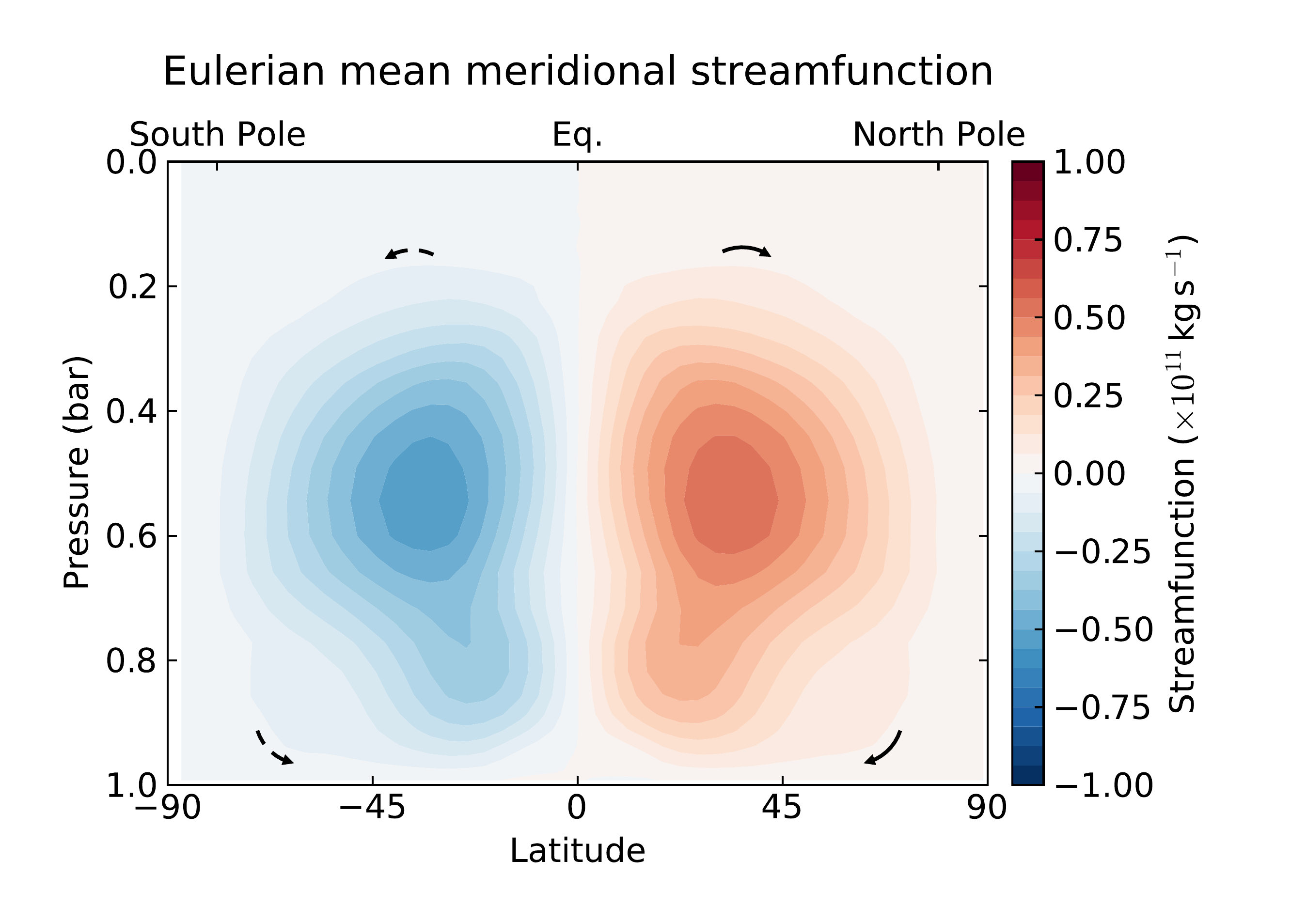}
    \caption{The overturning circulation of the terrestrial planet. Left panel: Tidally locked streamfunction $\Psi^{\prime}$ defined by Eq. \ref{eq:psi}. Right panel: Regular Eulerian mean meridional streamfunction $\Psi$ . Mass flux follows lines of constant $\Psi^{\prime}$ or $\Psi$. The direction of overturning is indicated by arrows and the sign of the streamfunction. Positive values indicate clockwise motion, negative values indicate anti-clockwise motion. Note that the colour scale differs in magnitude between the two panels. The maximum mass flux captured by $\Psi^{\prime}$ is much greater than that implied by $\Psi$, indicating that $\Psi^{\prime}$ captures more of the overturning than $\Psi$.}
    \label{fig:overturning}
\end{figure*}

$\Psi^{\prime}$ is shown in Fig. \ref{fig:overturning}. The overturning circulation viewed in the $\vartheta^{\prime}$--$p$ plane takes the form of a single overturning cell from the day-side to the night-side, and reveals each of the features of the circulation inferred from the $\lambda$--$\vartheta$ slices of $\boldsymbol{u}_{\text{d}}$ and $\omega$ which were shown in Fig. \ref{fig:div}. Air rises near the substellar point ($\vartheta^{\prime}=90^{\circ}$), before moving towards the night-side aloft. Descent begins as air passes across the terminator ($\vartheta^{\prime}=0^{\circ}$), and eventually air enters a near-surface return flow back towards the substellar point.

Visualising the overturning circulation in terms of $\Psi^{\prime}$ clearly captures the nature of the circulation better than the regular meridional mass streamfunction $\Psi=2\pi a\cos\vartheta\int^{0}_{p}[v]_{\lambda}\,\text{d}p/g$ (also shown in Fig. \ref{fig:overturning}), considered by previous authors \cite{merlis2010atmospheric,edson2011circulations,heng2011atmospheric,carone2015regimes,deitrick2020thor} The circulation strength indicated by $\Psi$ is far weaker than that indicated by $\Psi^{\prime}$ due to cancellation between day-side and night-side  meridional mass fluxes. The regular streamfunction $\Psi$ also captures none of the overturning in the zonal direction, or cross-polar meridional motion. We note that day-night differences in the overturning circulation have been described in previous works where $\Psi$ is averaged over a limited longitude range (i.e., over \emph{just} the day-side or night-side; see  \citealp{heng2011atmospheric, haqqmisra2018regimes}). However, this approach is difficult to motivate, as $\Psi$ is derived by integrating the continuity equation over \emph{all} longitudes. 

In summary, dividing the circulation of a terrestrial tidally locked atmosphere into rotational and divergent components corresponds to a division into two physically meaningful circulations. The rotational circulation is primarily composed of the stationary Rossby waves forced by the divergent circulation, and the zonal-mean jet produced by the stationary part of the atmospheric circulation. These two features can be further separated out by dividing the rotational circulation into contributions from the zonal-mean (the jet) and eddies (the stationary waves). The divergent circulation  corresponds to thermally direct overturning circulation that rises at the substellar point, diverges roughly isotropically, before subsiding on the night side and returning to the substellar point.

\subsection*{Hot Jupiter}\label{sec:hj}

\begin{figure*}[!t]
    \centering
    \includegraphics[width=.45\linewidth]{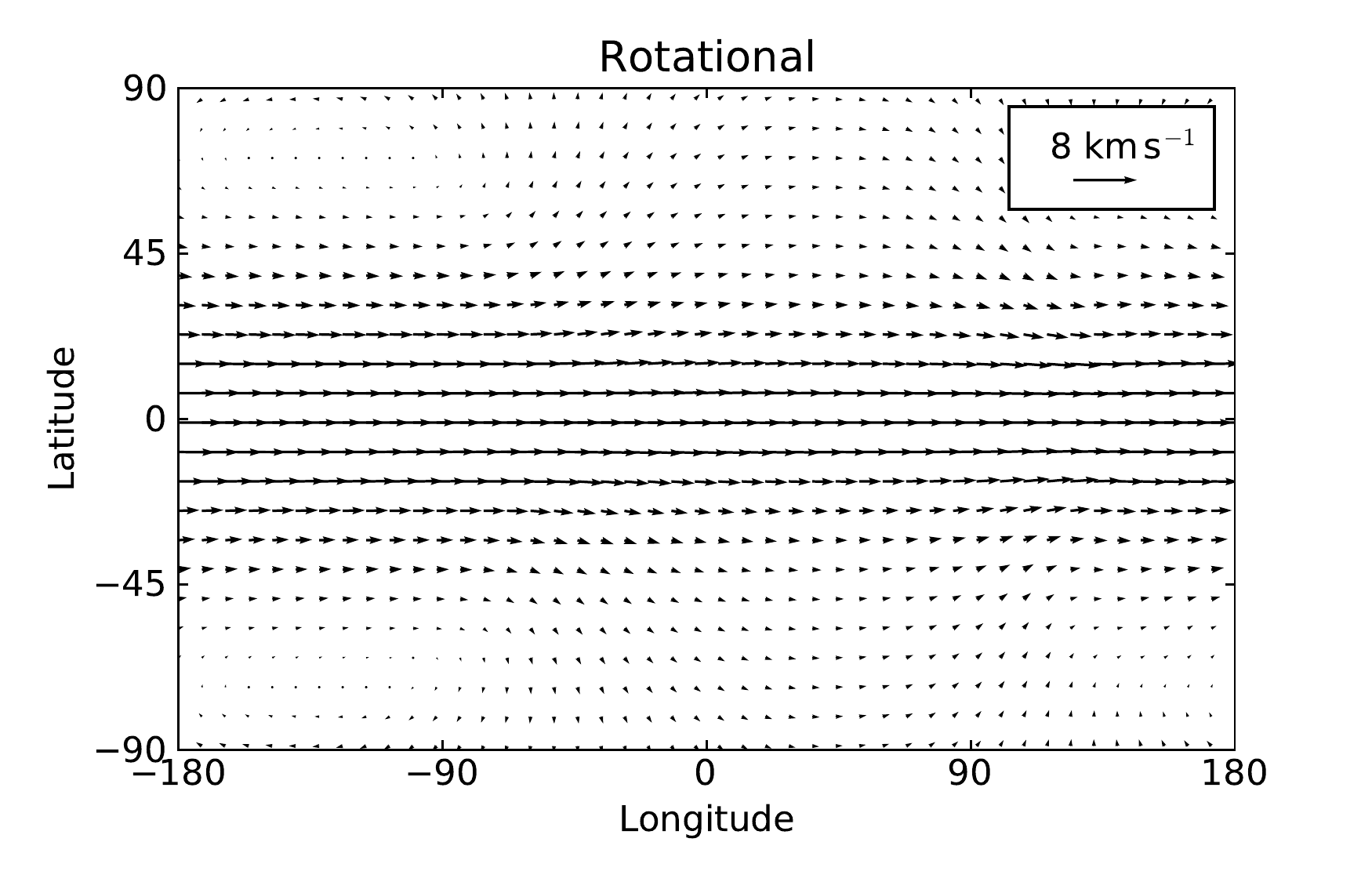}\hspace*{.35in}
    \includegraphics[width=.45\linewidth]{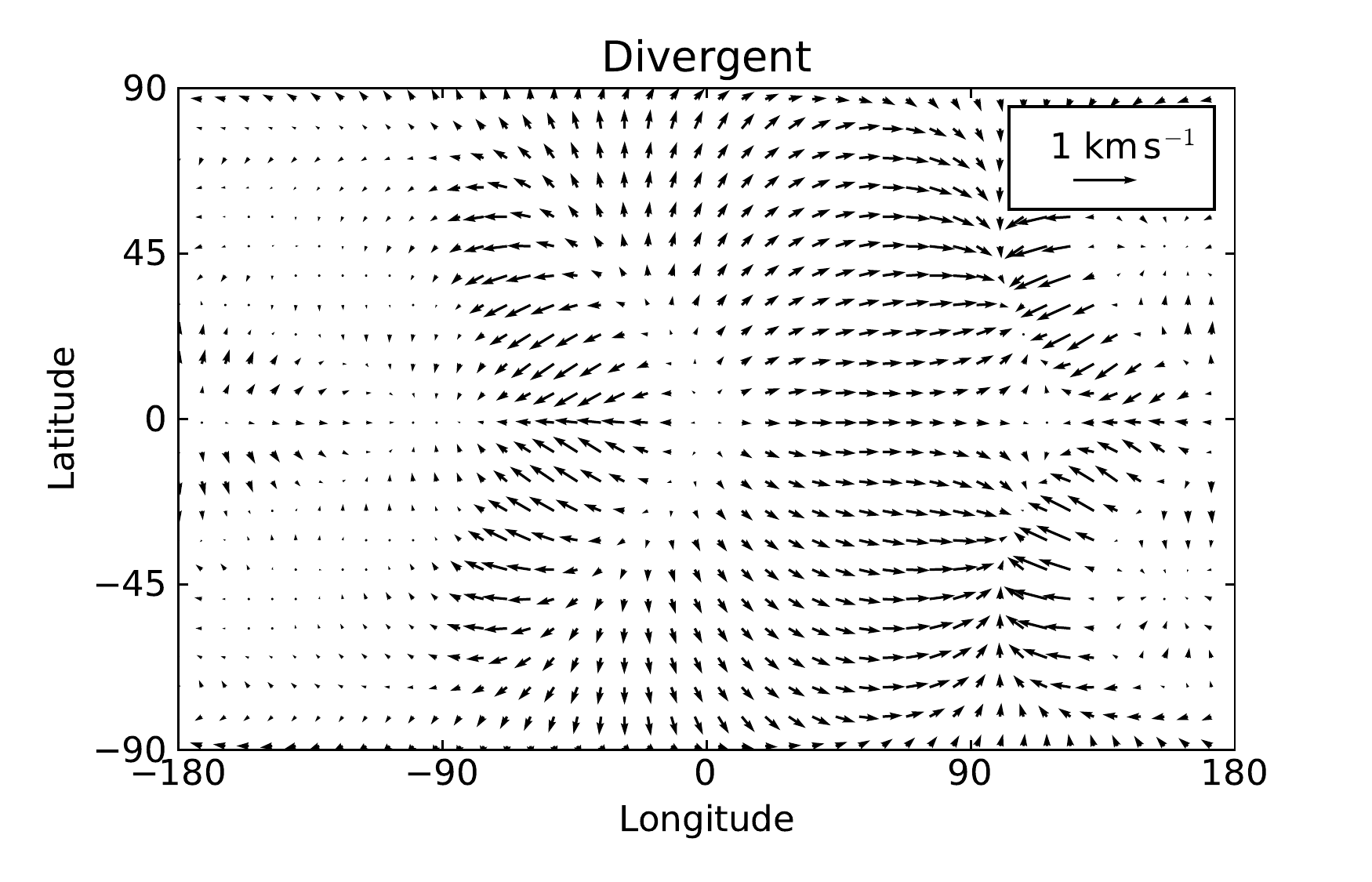}
    \caption{Helmholtz decomposition of horizontal velocity $\boldsymbol{u}$ for the hot Jupiter at 0.02 bar. Left panel: Rotational component of $\boldsymbol{u}$, $\boldsymbol{u}_{\text{r}}$, showing the eastward jet with a relatively weak stationary wave. Right panel: Divergent component of $\boldsymbol{u}$, $\boldsymbol{u}_{\text{d}}$, dominated by a divergent flow away from the substellar point ($0^{\circ},0^{\circ}$) which is less isotropic than the equivalent flow in the terrestrial simulation, which we suggest is due to the relatively stronger equatorial jet in the hot Jupiter atmosphere.}
    \label{fig:HJ-rot-div}
\end{figure*}

\begin{figure}[!t]
    \centering
    \includegraphics[width=0.90\linewidth]{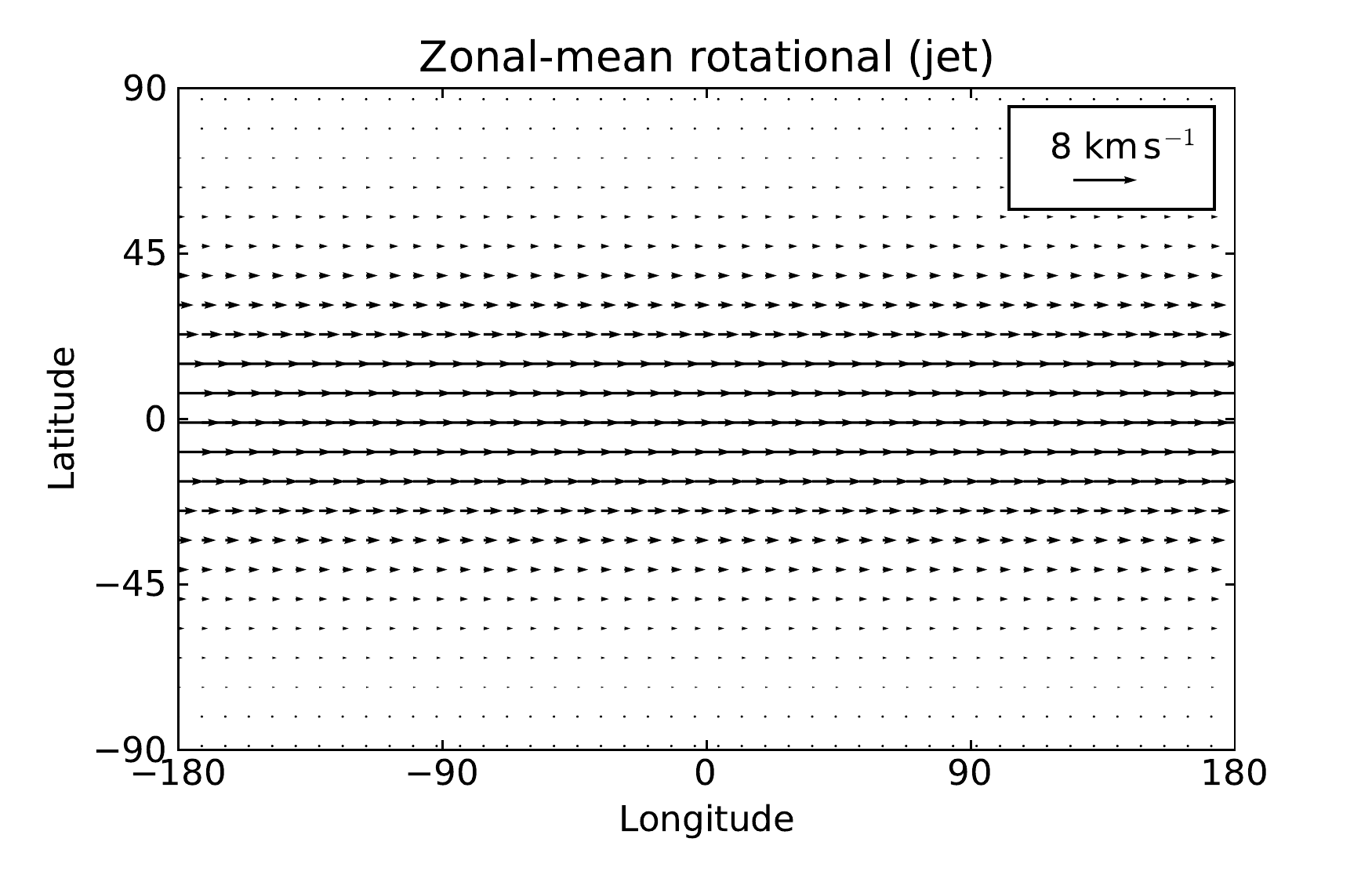} \\ \includegraphics[width=0.90\linewidth]{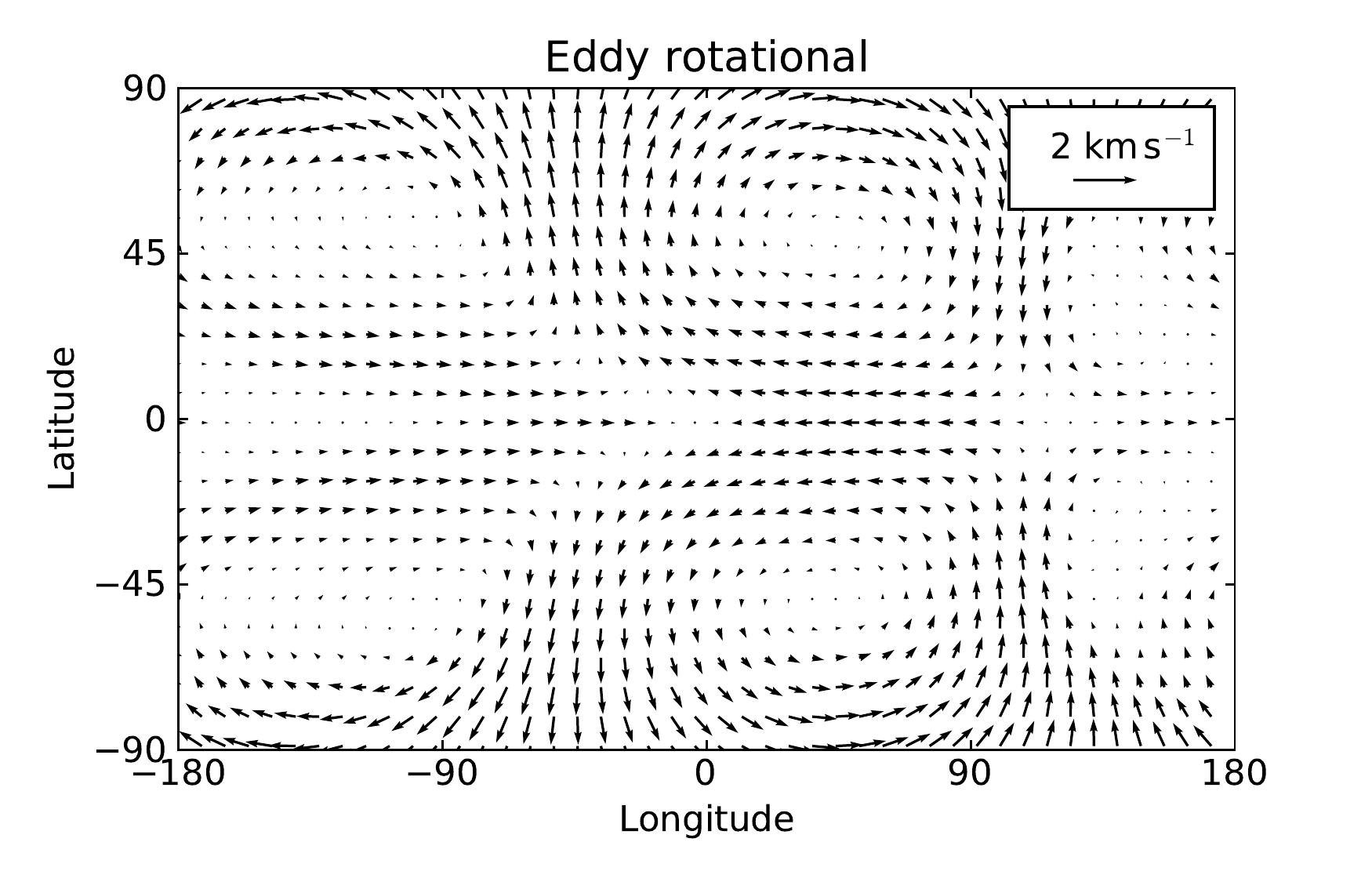}
    \caption{The two main parts of the rotational circulation shown in Fig. \ref{fig:HJ-rot-div}, decomposed as in Fig. \ref{fig:rot}. Top panel: Zonal-mean part of $\boldsymbol{u}_{\text{r}}$, $\overline{\boldsymbol{u}}_{\text{r}}$, showing the equatorial jet. Bottom panel: Eddy (total minus zonal mean) part of $\boldsymbol{u}_{\text{r}}$, $\boldsymbol{u}_{\text{r}}^{\prime}$, showing a wavenumber-1 wave that is weaker than the equatorial jet.}
    \label{fig:hj-rot}
\end{figure}

Hot Jupiters are gaseous tidally locked planets with radii comparable to Jupiter. They orbit close to their host stars, giving them orbital periods of a few days and temperatures above 1000 K. Their large sizes, short orbits, and high temperatures make them excellent targets for observing phase curves or measuring spectra. Understanding their global circulation and its effect on the transport of heat, chemical species, and clouds is vital to interpreting these observations \cite{crossfield2015observations,parmentier2017handbook}. In this section, we apply the Helmholtz decomposition (Eqs. \ref{eq:decomp}--\ref{eq:psi}) to a simulation of the hot Jupiter HD 189733b \cite{bouchy2005hd189773b} and show that it reveals the same circulation components that were in the terrestrial simulation.

Fig. \ref{fig:terr-circulation} shows the global circulation of this hot Jupiter, with the eastward equatorial jet and hot-spot shift typical of these planets. Fig. \ref{fig:HJ-rot-div} shows the decomposition of the total velocity field into the rotational and divergent velocity components $\boldsymbol{u}_{r}$ and $\boldsymbol{u}_{d}$, at the pressure level $0.02\,\text{bar}$, and Fig. \ref{fig:hj-rot} shows the further decomposition of the rotational component into contributions from the zonal mean and eddies. As with the terrestrial case, the Helmholtz decomposition shows that the total circulation is the sum of a divergent flow from day-side to night-side, plus a rotational flow dominated by a zonal-mean zonal jet and a wavenumber-1 stationary wave pattern. The divergent flow away from the substellar point is roughly isotropic in the north, south and eastward directions, but interaction with the zonal jet appears to prevent the divergent circulation from extending very far in the westward direction against the flow of the jet. The eddy component of the rotational flow is difficult to see in Fig. \ref{fig:HJ-rot-div}, as it is weaker than the jet, and it is more clearly appreciated in Fig. \ref{fig:hj-rot}. While the divergent circulation and eddy rotational circulation are weaker than the jet, neither is negligibly small, with characteristic velocities only a single order of magnitude less than the jet speed.

The eddy rotational component is dominated by a stationary wave response with zonal wavenumber 1, as it was in the terrestrial case, matching the assumptions of previous studies using shallow-water models \cite{showman2011superrotation,tsai2014three,hammond2018wavemean}. It also shows a clear `tilt' similar to that suggested by Showman and Polvani \cite{showman2011superrotation} to accelerate the equatorial jet, although the lack of a linear Rayleigh drag in the simulation means that this cannot be due to the linear Kelvin waves demonstrated by Showman and Polvani \cite{showman2011superrotation} to produce this tilt and acceleration. Hammond et al. \cite{hammond2020equatorial} showed how non-linear equatorial waves can produce this acceleration without an imposed linear drag, which we suggest is the process occurring in these GCM simulations.

\begin{figure*}[!t]
    \centering
    \includegraphics[width=.45\linewidth]{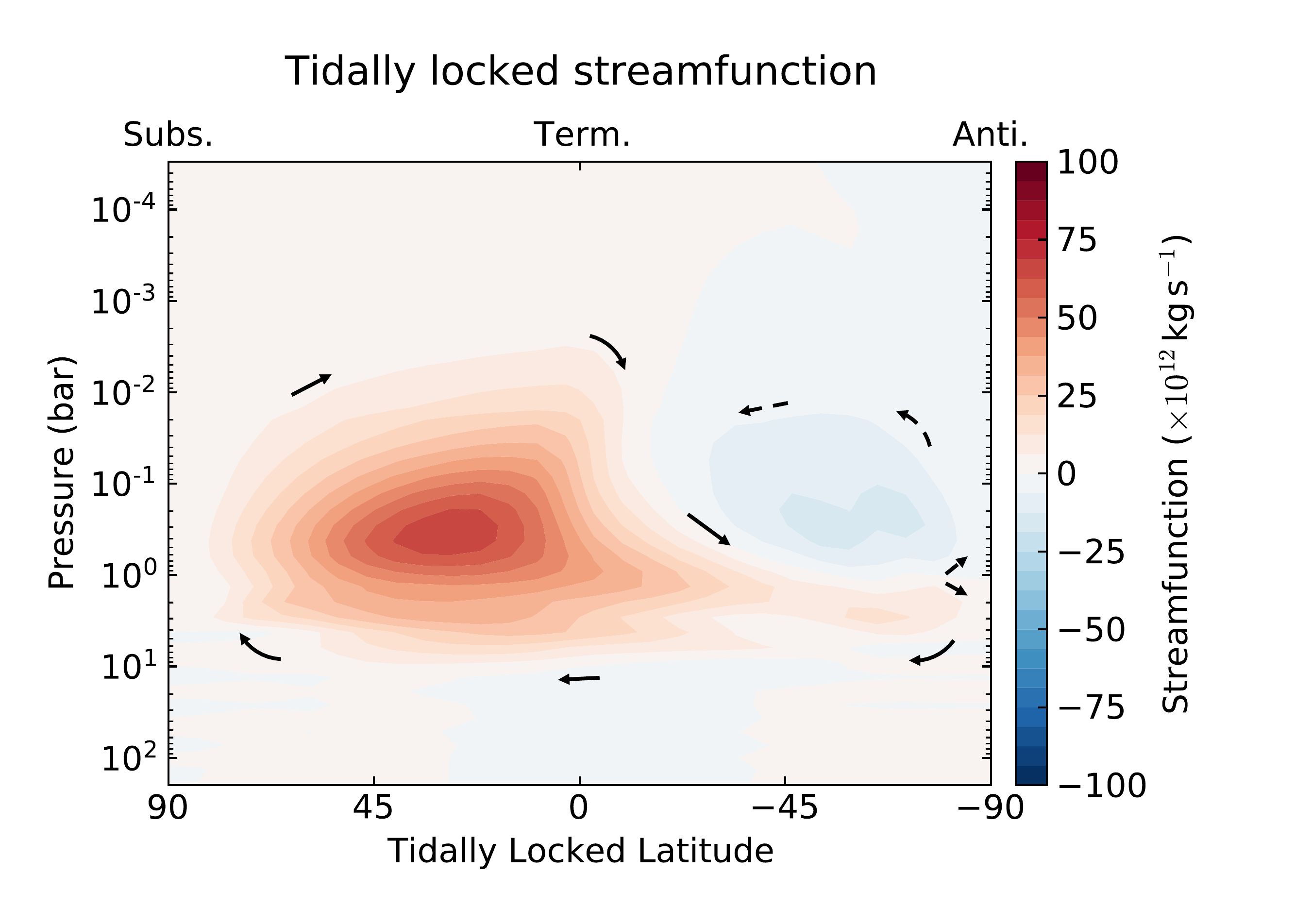}
    \includegraphics[width=.45\linewidth]{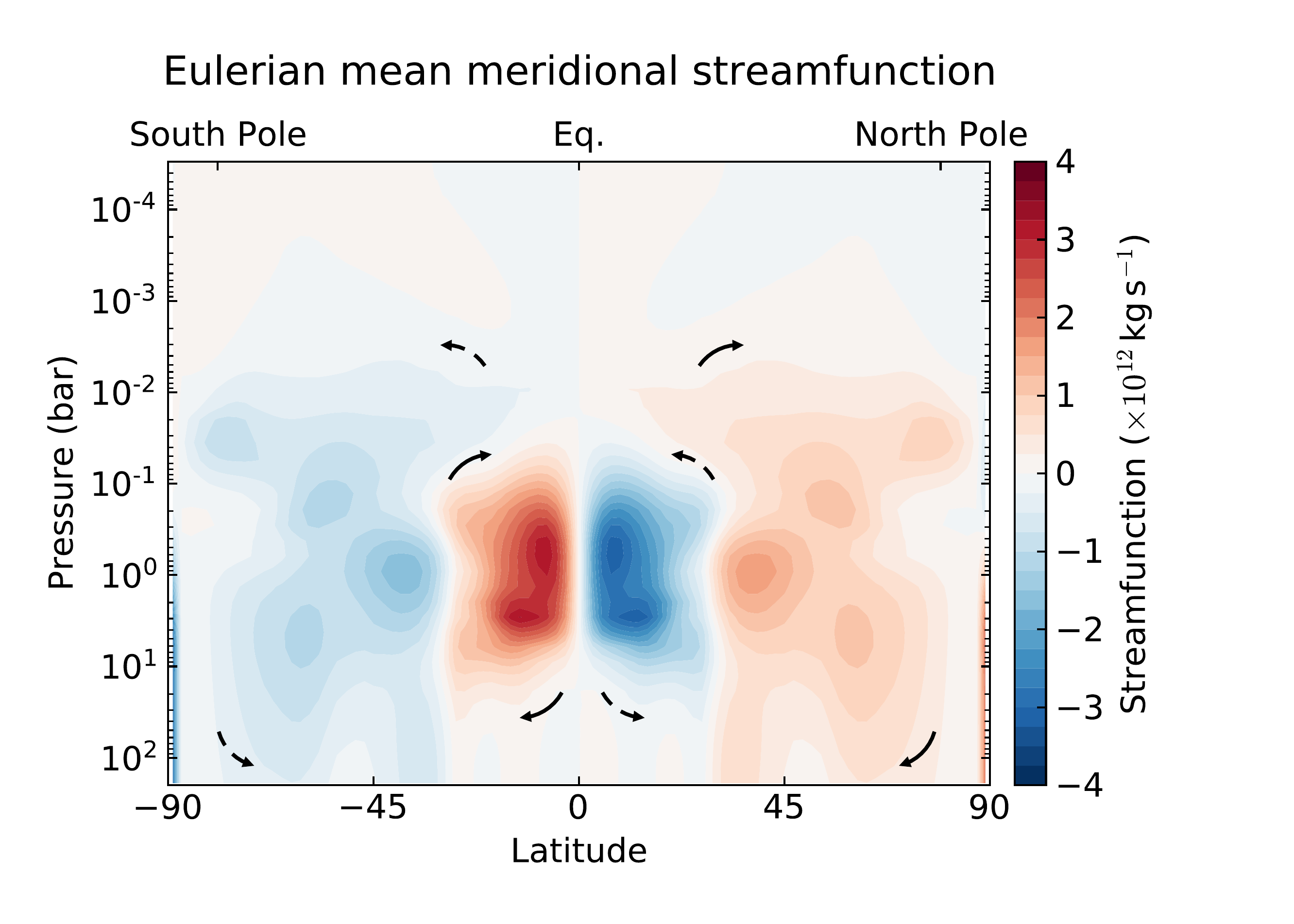}
    \caption{The overturning circulation of the hot Jupiter. Left panel: Tidally locked streamfunction $\Psi^{\prime}$ defined by Eq. \ref{eq:psi}. Right panel: Regular Eulerian mean meridional streamfunction $\Psi$ . Mass flux follows lines of constant $\Psi^{\prime}$ or $\Psi$. The direction of overturning is indicated by arrows and the sign of the streamfunction. Positive values indicate clockwise motion, negative values indicate anti-clockwise motion. As in the terrestrial case, $\Psi^{\prime}$ is much stronger than $\Psi$, showing that it captures much more of the overturning.}
    \label{fig:hj-overturning}
\end{figure*}

The tidally locked meridional mass streamfunction $\Psi^{\prime}$ (Eq. \ref{eq:sf}) is shown in Fig. \ref{fig:hj-overturning} to elucidate the vertical structure of the overturning circulation in the atmosphere of the hot Jupiter. Similarly to the terrestrial atmosphere, the main overturning circulation takes the form of a thermally direct cell rising on the day-side and sinking on the night-side just beyond the terminator. In addition to the main `clockwise' overturning cell, there is a weaker `anti-clockwise' cell on the night-side. At first glance, this circulation would appear to be thermally indirect. However, there is a local temperature maximum around $\vartheta^{\prime}=-90^{\circ}$, $p=1\,\text{bar}$ (see, e.g., the equatorial temperature profiles in Fig. 5 of  ref. \citealp{beltz2021hires}), generated by heat deposition by the extended `tongue' of the main clockwise circulation, which may drive the anti-clockwise circulation making it thermally direct.  Overturning motion in the hot Jupiter atmosphere is restricted to pressures less than roughly $3\,\text{bar}$. This is because there is little shortwave heating below this level to drive the circulation, as the shortwave optical depth is roughly $10$ at $3\,\text{bar}$ (see Table \ref{tab:sim-params}). Below $3\,\text{bar}$ the divergent circulation is quiescent (as is the zonally-averaged circulation shown in Fig. \ref{fig:terr-circulation}). Each of these features is consistent with the equatorial vertical velocity slices shown in Komacek et al. \cite{komacek2019vertical} for simulations with weak drag, which show multiple regions of upwelling and downwelling in longitude, concentrated above a pressure of approximately $1\,\text{bar}$. We note that while the main clockwise circulation is roughly isotropic, the weaker anti-clockwise circulation is concentrated to the west of the sub-stellar point. Future work investigating this feature could examine the three-dimensional structure of the overturning on the night-side, facilitated by isolation of the divergent circulation using the Helmholtz decomposition. It would also be of interest to study the overturning in terms of a three-dimensional residual mean flow such as that outlined in \cite{kinoshita20133dtem}.

The regular meridional mass streamfunction $\Psi$ is shown alongside $\Psi^{\prime}$ in Fig. \ref{fig:hj-overturning} for comparison. The divergent circulation has very little zonal symmetry, and consequently, the regular streamfunction (obtained by averaging over longitude) implies the existence of a weak `anti-Hadley' circulation, and an even weaker thermally-direct circulation in the mid-latitudes, that cannot be identified in the $\lambda-\vartheta$ slices of $\boldsymbol{u}_{d}$ shown in Fig. \ref{fig:HJ-rot-div}. $\Psi$ is approximately twenty times weaker than than $\Psi^{\prime}$, showing that the traditional streamfunction does not capture the majority of overturning in the simulation. This is mostly due to the cancellation of day-side and the night-side portions of the overturning cell in the zonal average. As for the terrestrial case, a streamfunction in latitude-longitude coordinates is not a good metric of the overturning circulation on tidally locked planets, as it misses the majority of the mass flux and implies the existence of an `anti-Hadley' circulation on the night-side \cite{heng2011atmospheric,charnay20153d,zhang2018globaltl,deitrick2020thor,mendonca2020heatmom} which is actually just the night-side portion of the single day-night overturning cell.

Comparing $\boldsymbol{u}_{d}$ and $\boldsymbol{u}_{r}$ for the terrestrial and hot Jupiter cases shows that the structure of these two components is very similar at the pressure level where the jet is strongest. This is notable, as the parameters of the hot Jupiter -- its composition, rotation rate, radius, temperature, and so on -- are very different to the terrestrial planet. This similarity indicates that a divergent circulation from the day-side to the night-side is a ubiquitous feature of tidally locked atmospheric circulations. This is in contrast with previous work \cite{merlis2010atmospheric,leconte2013tidallylocked,carone2015regimes,koll2015phasecurves,koll2016temperature,noda2017circulation} which has suggested that day-night thermally driven overturning is only an important feature of the circulation of slowly rotating tidally locked terrestrial planets.

\subsection*{Day-night heat transport}

The instellation gradient between the day- and night-sides of tidally locked planets naturally produces a day-night heat transport by the atmospheric circulation. This heat transport is a key feature of these atmospheres, due to its effect on atmospheric collapse and climate stability \cite{joshi1997tidally,heng2012collapse,wordsworth2015collapse,turbet2018modeling}, and observations of thermal phase curves \cite{schwartz2017phase}. In this section,
we assess the relative contribution of the rotational and divergent components of the circulation to heat transport from the day-side to night-side.

\begin{figure*}[!t]
    \centering
    \includegraphics[width=.45\linewidth]{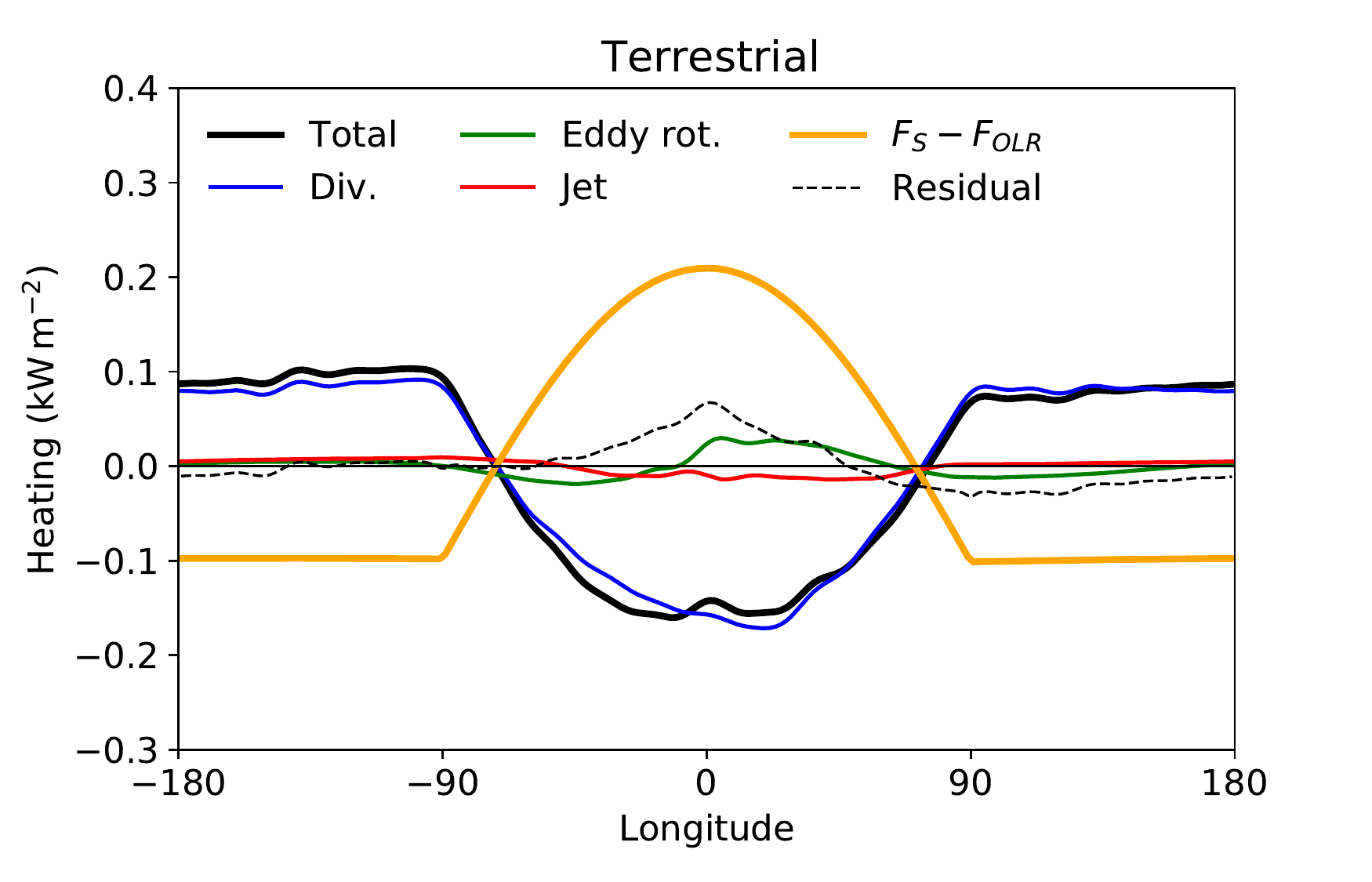}\includegraphics[width=.45\linewidth]{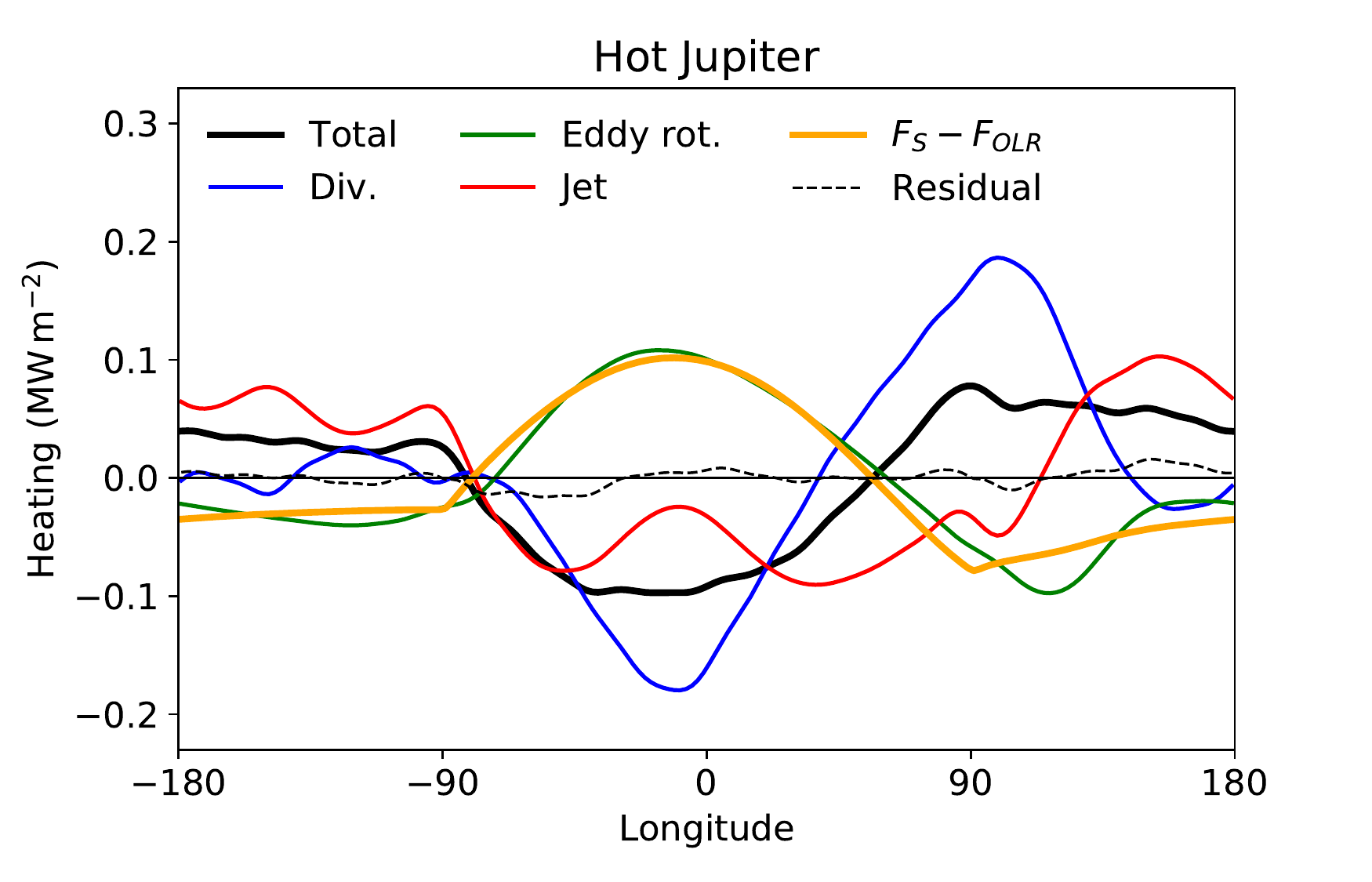}
    \caption{The dry static energy budget for the terrestrial planet and hot Jupiter. First panel: Terrestrial planet, showing that the day-night heat transport is dominated by the divergent (overturning) circulation, and the jet has little effect. Second panel: Hot Jupiter, showing that the jet and divergent circulation contribute roughly equally to day-night heat transport. The net effect of the rotational circulation on day-night heat transport is small, however, as the eddy rotational component transports heat from the night-side to the day-side, counteracting the jet.}
    \label{fig:dse-budget}
\end{figure*}

The atmospheric circulation transports heat by transporting dry static energy, which is
\begin{linenomath*}
\begin{equation} 
    s = c_{p}T + g z
\end{equation}
\end{linenomath*}
where $c_{p}$ is the heat capacity at constant pressure, $T$ is temperature, $g$ is acceleration due to gravity, and $z$ is height. Local heating is due to divergence of the dry static energy flux, and, column-integrated, the local energy budget is 
\begin{linenomath*}
\begin{equation}\label{eqn:local-dse-budget}
    \langle \nabla \cdot s\mathbf{u} \rangle + F_{S} - F_{OLR} = 0
\end{equation}
\end{linenomath*}
where $\langle \cdot \rangle$ denotes a vertical (mass-weighted) integration, $F_{S}$ is the net flux coming into the atmosphere from the stellar flux and the surface, and $F_{OLR}$ is the outgoing longwave radiation. We take a meridional average of Eq. \ref{eqn:local-dse-budget} (along lines of conventional latitude) and plot the terms in Fig. \ref{fig:dse-budget}, for the terrestrial planet and for the hot Jupiter. Merlis and Schneider \cite{merlis2010atmospheric} and Sergeev et al. \cite{sergeev2020atmospheric} calculated similar budgets for the moist static energy for tidally locked planets, but did not decompose it into contributions from the rotational and divergent components of the circulation. This budget does not exactly sum to zero, as there is a small residual introduced by the regridding to a latitude-longitude grid. It is possible to introduce a correction for this residual \cite{hill2017moist} but we have chosen to present the raw data for simplicity.

In the terrestrial case, the mean day-side instellation is 315 Wm$^{-2}$, of which 70\% is re-radiated from the day-side and 26\% is transported to the night-side. Of the transport to the night-side, 96\% is by the divergent circulation $\mathbf{u}_{d}$, 6\% is by the zonal mean jet part of the rotational circulation, and $-2$\% is by the eddy part of the rotational circulation (back from the night-side to the day-side). This is a surprising result, as Fig. \ref{fig:decomp} shows that the divergent circulation is significantly weaker than the rotational circulation in terms of absolute velocity, yet it transports far more energy to the night-side. Divergent heat transport is roughly symmetric in the eastward and westward directions in the terrestrial simulation, reflecting the isotropic nature of the divergent circulation.

The dominance of the divergent circulation in this case can be explained by the fact that the atmosphere is in a `Weak Temperature Gradient' (WTG) regime \cite{mills2013utility,pierrehumbert2018review}. In this regime, which is typical of planets that are slowly rotating, the zonal momentum equation is dominated by a balance between the local temperature gradient and a term that depends on the square of the local velocity (rather than the Coriolis force; see \citealp{hammond2020equatorial}). This means that the temperature gradients are relatively small. Expanding the local heating due to the divergence of dry static energy flux
\begin{linenomath*}
\begin{equation}
    \nabla \cdot s\mathbf{u} = s \nabla \cdot \mathbf{u} + \mathbf{u} \cdot \nabla s \label{eq:expand-dse}
\end{equation}
\end{linenomath*}
shows that the contribution of the rotational velocity $\mathbf{u}_{r}$ will only be through the second term on the right-hand-side, as $\nabla \cdot \mathbf{u}_{r}=0$. However, the horizontal gradient of the dry static energy $\nabla s$ is small in the WTG regime, so the heating due to the $\boldsymbol{u}\cdot\nabla s$ terms is small, meaning that the total heating due to the rotational circulation is small. On the other hand, the divergent velocity $\mathbf{u}_{d}$ has non-zero divergence so contributes to the local heating via both terms on the right-hand-side of Eq. \ref{eq:expand-dse}. This means that the divergent circulation plays a dominant role in the energy budget, as $s\nabla\cdot\boldsymbol{u}_d$ is much greater than $\boldsymbol{u}_r\cdot\nabla s$ (or $\boldsymbol{u}_d\cdot\nabla s$).

For the hot Jupiter, the mean day-side instellation is 119375 Wm$^{-2}$, of which 70\% is re-radiated from the day-side and 35\% is transported to the night-side by the total circulation. Note that the total residual is larger in the calculation of the dry static energy budget of the hot Jupiter than for the terrestrial case, so these percentages do not add up to exactly 100\%. 78\% of the day-night heat transport is by the divergent circulation and 22\% by the rotational circulation. The rotational circulation only makes a small contribution because its two parts almost cancel each other out. The zonal-mean jet transports 122\% of the eventual net day-night transport to the night-side, but the eddy part of the rotational circulation moves $-100$\% of the net transport back towards the day-side. The near-cancellation of these two processes results in the rotational circulation only contributing 22\% of the total day-night heat transport.

The hot Jupiter is less firmly in the WTG regime as it is more strongly forced, rotating faster, and larger than the terrestrial planet \cite{pierrehumbert2018review}. Additionally, the divergent circulation on the hot Jupiter is less effective at transporting heat than on the terrestrial planet, as it cannot extend very far away from the substellar point to the west, due to interaction with the stronger zonal mean jet (note that divergent heat transport is not symmetric about $0^{\circ}$ longitude, unlike in the terrestrial atmosphere). Both of these features lead to stronger day-night gradients in $s$ than are found in the terrestrial atmosphere, which enhances rotational heat transport in the hot Jupiter atmosphere by making the second term in Eq. \ref{eq:expand-dse} larger. This leads to enhanced heat transport from the day-side to the night-side by the superrotating jet, as the jet flows through a negative gradient in $s$ (temperature decreasing relative to flow direction) downstream of the substellar point. The eddy rotational circulation flows back towards the substellar point near the equator (see Fig. \ref{fig:hj-rot}) so partially counteracts day-night heat transport by the jet by returning heat to the day-side. The net effect is that the majority of day-night heat transport on the hot Jupiter is due to the divergent circulation. Unlike the terrestrial case, the rotational circulation also provides a first-order contribution to the heat budget due to the larger gradients in dry static energy on the hot Jupiter. 

The fact that the majority of day-night heat transport is due to the divergent circulation, for both the terrestrial planet and the hot Jupiter, is an important result, as some previous work studying day-night heat transport on tidally locked planets has focused entirely on the role of the jet and stationary waves  \cite{perez2013atmospheric,hammond2017climate,mendonca2020heatmom}. The red line in Fig. \ref{fig:dse-budget} (minus the known instellation) corresponds exactly to the broadband thermal phase curve that could be observed for the planet \cite{parmentier2017handbook}, and consequently the energy budget can be constrained by observations. Note that the line in Fig. \ref{fig:dse-budget} would be flipped horizontally when viewed as a phase curve, as it is plotted as a function of longitude rather than orbital phase here.  Phase curves observed from hot Jupiters are already available and will increase in quality in the coming years \cite{bean2018transiting}. Similar observations should become available for terrestrial tidally locked planets when next-generation telescopes such as the \emph{James Webb Space Telescope} come online \cite{deming2009jwst,koll2015phasecurves}. Separating the effects of the rotational and divergent circulations to heat transport on tidally locked planets over a broad range of parameters could be crucial to interpreting phase curve observations.

\section*{Discussion}

\begin{figure}
    \centering
    \includegraphics[width=.95\linewidth]{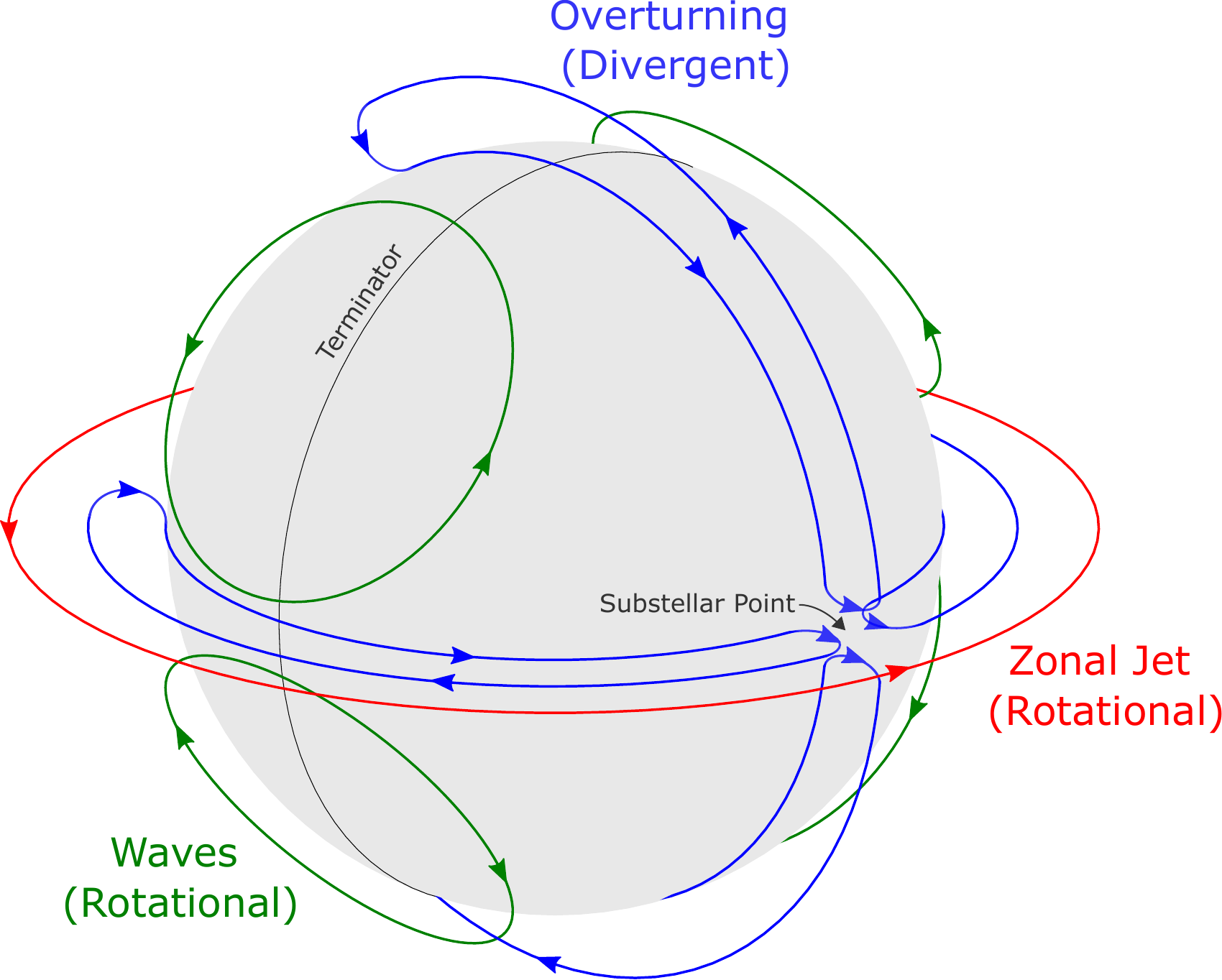} 
    \caption{A schematic showing the three main circulation components identified in this study. Divergent, overturning circulation (blue) rises at the substellar point and extends roughly isotropically across the terminator, before descending on the night-side at a location that depends on the strength of the circulation. The rotational circulation is divided into the zonal-mean jet (red) and the eddy stationary waves (green). The divergent circulation has approximate symmetry around the axis connecting the substellar and antistellar points, which motivates the use of the tidally locked coordinate system to analyse it.}
    \label{fig:summary-diagram}
\end{figure}

\subsection*{Summary} In this work, we have applied a Helmholtz decomposition (Eq. \ref{eq:decomp}) to the circulation of two well-studied benchmark simulations of tidally locked planets. We considered the case of a terrestrial planet using the Exo-FMS GCM \cite{hammond2018wavemean,hammond2020equatorial}, designed to model planets similar to those in the Trappist-1 system \cite{gillon2017seven}, and a hot Jupiter using the THOR model \cite{mendoncca2016thor,deitrick2020thor}, configured with the parameters of the planet HD 189733b \cite{bouchy2005hd189773b}.

Our work has three main conclusions.  First, applying a Helmholtz decomposition to the atmospheric circulation of tidally locked planets isolates the three main components that make up the total circulation -- the superrotating jet, stationary waves, and overturning circulation. Fig. \ref{fig:summary-diagram} shows the qualitative form of each of these three components. The rotational circulation contains the jet (zonal-mean rotational) and wavenumber-1 stationary wave (eddy rotational), and the divergent component contains the overturning circulation. The stationary wave corresponds to the shallow-water waves analysed by previous studies \cite{showman2011superrotation,tsai2014three,hammond2018wavemean}. For both cases, the divergent circulation is weaker than the rotational circulation, but it is not negligible, with velocities only a single order of magnitude less than those associated with the rotational circulation. The overturning circulation is similar in some respects to a combination of the Hadley and Walker circulations on Earth. However, it occurs on a much larger scale (the planetary scale), and its isotropy means that there is little distinction between the overturning in the zonal or meridional directions. With this in mind, we suggest it is not useful and potentially misleading to define distinct `Hadley' and `Walker' circulations for tidally locked planets.

Second, the overturning circulation is represented naturally by a mass streamfunction in tidally locked coordinates \cite{koll2015phasecurves}, as it is dominated by isotropic motion away from the substellar point. This reveals that the overturning circulation is dominated by a single cell which features air rising near the substellar point and descending on the night-side. By contrast, the traditional Eulerian mean meridional mass streamfunction defined in latitude-longitude coordinates fails to capture most of the overturning, and contains no information about cross-polar flow or overturning in the zonal direction. 

Third, the divergent circulation associated with the day-night overturning dominates the day-night heat transport even when its absolute velocity is weaker than the other components. For the terrestrial case, this is because these planets are typically in a `Weak Temperature Gradient' regime, which inhibits advection of heat by the superrotating jet and stationary waves. For the hot Jupiter, the jet and stationary waves do make a significant contribution to the heat budget. However, the direction of heat transport associated with each component is opposite from one another. The jet transports heat from the day-side to the night-side, but the stationary waves transport heat back from the night-side to the day-side; these opposing transports lead to a large cancellation, meaning that the divergent circulation still dominates the net day-night heat transport. The dominance of the divergent circulation in the day-night heat budget on both the terrestrial planet and the hot Jupiter illustrates that each of the circulation components is important to the whole circulation, even in regimes where one component has much higher velocities than the others.

\subsection*{Future work} We now outline a few opportunities for future research motivated by this study. First, in this study we have only considered two `exemplar' benchmark simulations of tidally locked planets. In the future, the division of the global circulation into its rotational and divergent components should be applied to simulations spanning a wide parameter space. Such a study could investigate how the strength of each circulation component depends on planetary parameters such as the instellation, radius, rotation rate, and planetary mass. In addition, simulations of the circulation on both terrestrial planets and hot Jupiters have been shown to be sensitive to choices made in model configuration. For example, the circulation of a hot Jupiter has been shown to be sensitive to imposed drag and radiative timescales \cite{komacek2016daynighti}, and the lower boundary pressure \cite{carone2020wasp}. Inspection of the \emph{ultra-hot} Jupiter simulated in Tan and Showman \cite{tan2019atmospheric} suggests a stronger divergent circulation than the hot Jupiter we show above, possibly due to its shorter radiative timescale. On terrestrial tidally locked planets, the presence or absence of a land surface (as opposed to a homogeneous ocean surface) has been shown to affect the atmospheric circulation \cite{lewis2018influence}. Additionally, on habitable planets, water vapour and clouds will likely influence the atmospheric dynamics and heat transport, through both their radiative effects and the effect of latent heating \cite{merlis2010atmospheric}. The terrestrial simulation we have analysed in the present work is dry (i.e., there is no water vapour or cloud condensate in the model atmosphere), and examining the effect of moisture on the Helmholtz decomposition of the circulation and heat transport is an obvious follow-up to this work. Understanding what determines the relative strength of the rotational and divergent circulation components on different classes of tidally locked planets is important to determining when one will dominate the other, which would help ascertain which processes are most important for transporting heat and shaping thermal phase curve observations in different parameter regimes.

Second, a study of a large parameter space of simulations would also be a basis for testing simple models of the different components of the circulation that we have identified. Hammond et al. \cite{hammond2020equatorial} developed a simple model to predict the speed of the equatorial jet on terrestrial tidally locked planets, which is the zonal-mean part of the rotational circulation that we analysed above. This helps with predicting the hot spot shift seen in thermal phase curves. Developing models that describe the divergent circulation would allow predictions of day-night heat transport and would aid interpretation of the amplitude of thermal phase curves. Models of the Earth's Walker circulation such as that of Bretherton and Sobel \cite{bretherton2002walker} would be a good starting point for a simple model of the overturning circulation on tidally locked planets, as we have shown that the divergent circulation is approximately isotropic in all directions from the substellar point which means that it is only weakly influenced by the effect of the planet's rotation. Such a model would be particularly useful for the slowly rotating terrestrial case \cite{koll2015phasecurves}, where the day-night heat transport is dominated by the divergent circulation so that a model of the divergent circulation could predict the total heat transport and the resulting broadband thermal phase curve. When the jet, stationary waves, and overturning circulation are comparably important to the heat budget (as was the case for the hot Jupiter) one could fit a phase curve to a sum of three basis functions which represent the shape of each component. Phase curves have previously been fitted to a sum of spherical harmonics \cite{rauscher2018more,luger2019starry} that do not correspond to physical components of the circulation.

Finally, understanding the strength of the divergent circulation (which accounts for all vertical motion) is vital to models and observations of disequilibrium chemistry and cloud formation. These models often rely on a `$K_{zz}$' parameter that describes the strength of diffusive mixing (a parametrisation for vertical advection); deriving an estimate of this from a model of the divergent circulation would constrain an aspect of these models that is not well understood (see \citealp{zhang2018global} for various ways to estimate this parameter). Modelling the transport due to the rotational and divergent circulations of atmospheric tracers would also be valuable to understanding the distribution of clouds \cite{parmentier2016transitions}, the distribution of chemical species in disequilibrium models \cite{tsai2017vulcan}, and the heating due to recombination of dissociated hydrogen \cite{bell2018increased}.

This study has shown how the global circulation of tidally locked planets can be divided into three physically meaningful components, summarised in Fig. \ref{fig:summary-diagram}. Each of these components plays an important role, even when they are not immediately apparent in the total circulation. We hope that this will form a basis for future work on the behaviour of each of these components and their role in the global circulation.

\matmethods{
\subsection*{Description of General Circulation Models}

The terrestrial simulation was run using `Exo-FMS' \cite{hammond2020equatorial}, and the hot Jupiter simulation was run using `THOR' \cite{deitrick2020thor}. We chose to use these existing simulations as benchmarks, as they have already had their basic circulations analysed in  \cite{hammond2020equatorial} and \cite{deitrick2020thor}; this allows us to progress directly to analysing the new circulation features revealed by the Helmholtz decomposition.

\begin{table*}%[tbhp]
\centering
\caption{Parameters of the terrestrial and gaseous simulations in the models Exo-FMS and THOR. Both models use semi-grey radiative transfer schemes and dry convective adustment. \cite{hammond2018wavemean} and \cite{deitrick2020thor} discuss the respective models and simulations in more detail. We have included columns with the parameters of Earth and Jupiter for comparison, for which we have quoted the non-dimensional lengthscales estimated in \cite{showman2010atmospheric}.}\label{tab:sim-params}
\begin{tabular}{lrrrr}
& \textbf{Terrestrial (FMS)} & \textbf{Gaseous (THOR)} & \textbf{Earth} & \textbf{Jupiter}  \\ 
\midrule
\textbf{Planetary parameters} & &  \\
%\midrule
Radius $a$ (m) & $6.371 \times 10^{6}$ & $7.970 \times 10^{8}$ & $6.371 \times 10^{6}$ & $7.149 \times 10^{8}$\\
Rotation rate $\Omega$ (s$^{-1}$) & $7.29 \times 10^{-6}$ & $3.279 \times 10^{-5}$ & $7.29 \times 10^{-5}$ & $1.76 \times 10^{-4}$  \\
Gravitational acceleration $g$ (ms$^{-2}$) & 9.81 &  21.4 & 9.81 & 24.79\\
Lower boundary pressure $p_{0}$ (Pa) & $10^{5}$ &  $2.2 \times 10^{7}$ & $10^{5}$ & - \\
Gas constant $R$ (JK$^{-1}$kg$^{-1}$) & 287 & 3779 & 287 & 3745\\
Heat capacity $c_{p}$ (JK$^{-1}$kg$^{-1}$) & 1005  &  13226.8 & 1005 & 12359.1 \\
Instellation $F_{S}$ (Wm$^{-2}$) & 1000 &  467072 & 1361 & 50.26\\
\midrule
\textbf{Radiative parameters} & &  \\
Albedo $A_{0}$ & 0 &  0.18 & - & -\\
Longwave optical depth scaling $n_{LW}$ & 1 & 2 & - & -\\
Longwave optical depth $\tau_{LW}$ & 1 &  4680 & - & -\\
Shortwave optical depth scaling $n_{SW}$ & - & 1 & - & - \\
Shortwave optical depth $\tau_{SW}$  & 0 &  1170 & - & - \\
\midrule 
\textbf{Approximate non-dimensional lengthscales} & & \\ 
%\midrule
Equatorial deformation scale $\mathcal{L}_{\text{eq}}/a$ & 1.77 & 0.68 & 0.3 & 0.03 \\ 
Rhines scale $\mathcal{L}_{\text{R}}/a$ & 0.63 & 0.82 & 0.5 & 0.1\\ 
\bottomrule
\end{tabular}
\end{table*}

\subsubsection*{Terrestrial planet} The terrestrial simulation was run in `Exo-FMS', which solves the primitive equations on a cubed-sphere grid with a vertical sigma-pressure coordinate \cite{hammond2020equatorial}. These equations are solved for a dry atmosphere with no representation of water vapour or cloud condensate. Exo-FMS was built on the cubed-sphere dynamical core of the `Flexible Modelling System' \cite{lin2004fv}. 

The simulation of a terrestrial planet in Exo-FMS was configured in exactly the same way as the simulation in \cite{hammond2020equatorial} with instellation 1000 Wm$^{-2}$. It uses semi-grey two-stream radiative transfer and dry convective adjustment (as in \citealp{hammond2018wavemean} and \citealp{hammond2017climate}), with atmospheric parameters listed in Table \ref{tab:sim-params}. We chose to use an albedo of zero to simplify the number of parameters chosen, as the value of the albedo is degenerate with the strength of the instellation. Each of the six faces of the cubed-sphere grid has 48 by 48 cells, which corresponds
approximately to a latitude-longitude resolution of $1.9^{\circ}$. The model has 48 vertical levels (this is the only difference to the simulation in \citealp{hammond2020equatorial}). The model is stabilised with a fourth-order hyperdiffusion with a coefficient the same as the default value used for Earth-like simulations distributed with this cubed-sphere dynamical core \cite{lin2004fv}. A linear drag is applied to the velocities at the surface with a timescale of 1 day, which decreases in strength linearly with pressure up to at value of zero at 700 mbar. The test was spun up for 1000 (Earth) days, until the total angular momentum and total outgoing longwave radiation were equilibrated, and the output data was taken as an average over the next 1000 days. Its orbital period of 10 days is an approximate representation of a tidally locked Earth-sized and Earth-temperature planet orbiting an M-dwarf, similar to the Trappist-1 planets \cite{gillon2017seven}.

\subsubsection*{Hot Jupiter} The simulation of the hot Jupiter was run in `THOR', which solves the nonhydrostatic Euler equations on an icosahedral grid with a vertical height coordinate \cite{deitrick2020thor}. The `deep' form of these equations is solved, meaning that the vertical coordinate $z$ is always used instead of the radius $r$, which is substituted in some parts of the `shallow' equations used in the simulation of the terrestrial planet in FMS. We are grateful to Russell Deitrick and the rest of the THOR development team who provided us with the data for this simulation.

The simulation of the hot Jupiter is exactly the same as the nonhydrostatic benchmark of HD 189733b presented in \cite{deitrick2020thor}. This simulation uses semi-grey radiative transfer and dry convective adjustment, just like the simulation in Exo-FMS, with parameters listed in Table \ref{tab:sim-params}. The icosahedral grid has a horizontal resolution corresponding approximately to $2^{\circ}$, and 40 vertical levels. The model is stabilised with a fourth-order hyperdiffusion as discussed in \cite{deitrick2020thor}, as well as a sponge layer applying a weak linear drag to the eddy part of the zonal velocity near the top of the atmosphere. The simulation was spun up for 9000 days and then output data was recorded as an average over the next 1000 days. \cite{deitrick2020thor} discusses the spin-up of this simulation. Their Fig. 29 shows that the total angular momentum has equilibrated after 5000 days. This is a typical hot Jupiter that has been well characterised observationally, and its atmospheric dynamics have been simulated and analysed in previous studies (see e.g. \citealp{showman2015circulation}).

\subsubsection*{Non-dimensional lengthscales} In addition to the dimensional parameters used to configure each model, Table \ref{tab:sim-params} shows two non-dimensional lengthscales: the equatorial deformation scale, and the Rhines scale, computed to facilitate future comparison between our two experiments, and those of other work. 

The equatorial deformation scale is defined 
\begin{linenomath*}
\begin{equation}
    \frac{\mathcal{L}_{\text{eq}}}{a}=\frac{1}{a}\left(\frac{c}{\beta}\right)^{\frac{1}{2}}, \label{eq:leq}
\end{equation}
\end{linenomath*}
where $c=\sqrt{gH}$ for the barotropic mode \cite{vallis2017atmospheric}, and $a$ is the planetary radius. $\beta=\text{d}f/\text{d}y$, where $f=2\Omega\sin\theta$, with $\Omega$ the planetary rotation rate, and $y=a\vartheta$. We approximate $\beta$ as $\beta=2\Omega/a$. When calculating $c$, we use $H=RT/g$ for the scale height where $R$ is the specific gas constant, $T$ is a characteristic temperature and $g$ is the acceleration due to gravity. We use the emission temperature as a characteristic temperature; our calculation is not sensitive to this choice, and similar values for $\mathcal{L}_{\text{eq}}/a$ are obtained if we use the globally-averaged temperature at the model bottom, for example. 

The non-dimensional Rhines length is defined 
\begin{linenomath*}
\begin{equation}
    \frac{\mathcal{L}_{\text{R}}}{a} = \frac{1}{a}\left(\frac{2U}{\beta}\right)^{\frac{1}{2}},\label{eq:rhines}
\end{equation}
\end{linenomath*}
where $U$ is a characteristic horizontal velocity scale (\citealp{rhines1975scale}; e.g., the root mean squared velocity). We use $U=16\,\text{ms}^{-1}$ for the terrestrial planet, and $U=1760\,\text{ms}^{-1}$ for the hot Jupiter, which are the root mean squared horizontal velocities at a pressure level characteristic of the jet height ($0.4\,\text{bar}$ for the terrestrial case, $0.02\,\text{bar}$ for the hot Jupiter).

\subsection*{Numerical solution of Helmholtz decomposition}

The rotational and divergent velocities $u_r$ and $u_d$ are obtained using the python library \texttt{windspharm} \cite{dawson2016windspharm}, using the routine \texttt{helmholtz}. We use an $n=21$ spherical harmonic truncation. Our results are insensitive to an increase in $n$, as long as $n$ is small enough that it is not sensitive to spurious features near the poles, introduced by interpolating from the model grids to a latitude-longitude grid.

\subsection*{Tidally locked coordinate}

Koll and Abbot \cite{koll2015phasecurves} propose the `tidally locked coordinate system' $\left(\vartheta^{\prime}, \lambda^{\prime}\right)$  as an alternative coordinate system to the traditional latitude-longitude system $(\vartheta, \lambda)$, which makes use of the symmetry present in the atmosphere of a tidally locked planet. The coordinates are effectively a rotation of regular latitude-longitude coordinates, so that the polar axis runs from the substellar point to the antistellar point. They define the tidally locked latitude $\vartheta^{\prime}$ to be the angle to the terminator, and the tidally locked longitude to be the angle about the substellar-antistellar axis. They choose $(\vartheta, \lambda)=(0,0)$ to be the substellar point and $\left(\vartheta^{\prime}, \lambda^{\prime}\right)=(0,0)$ to be the antistellar point. In this coordinate system, lines of constant tidally locked longitude run from the substellar point to the antistellar point. 

Koll and Abbot \cite{koll2015phasecurves} show that latitude-longitude coordinates can be transformed into tidally locked coordinates by the transformation
\begin{linenomath*}
\begin{align}
\vartheta^{\prime}&=\sin ^{-1}(\cos \vartheta \cos \lambda) \\
\lambda^{\prime}&=\tan ^{-1}\left(\frac{\sin \lambda}{\tan \vartheta}\right).
\end{align}
\end{linenomath*}
Scalar fields (e.g., $\omega$) in latitude-longitude coordinates are transformed by transforming their locations into tidally locked coordinates then interpolating onto a regular grid. The vector horizontal velocity $\boldsymbol{u}=(u, v)$ is transformed into $\boldsymbol{u}^{\prime}=(u^{\prime}, v^{\prime})$ using the relation
\begin{linenomath*}
\begin{align}
u^{\prime} &=\cos \vartheta^{\prime}\left(\frac{\partial \lambda^{\prime}}{\partial \lambda} \frac{u}{\cos \vartheta}+\frac{\partial \lambda^{\prime}}{\partial \vartheta} v\right), \\
v^{\prime} &=\frac{\partial \vartheta^{\prime}}{\partial \lambda} \frac{u}{\cos \vartheta}+\frac{\partial \vartheta^{\prime}}{\partial \vartheta} v .
\end{align}
\end{linenomath*}
derived in \cite{koll2015phasecurves}. We calculate these transformations using a code provided by D. Koll at \url{https://github.com/ddbkoll/tidally locked-coordinates}.

}
\showmatmethods{} % Display the Materials and Methods section

\acknow{We are grateful for the comments of two anonymous referees that helped us to refine this work throughout. We thank Russell Deitrick and all the developers of the THOR GCM for sharing the results of the simulation of HD 189733b with us, and for making their model publicly available. We thank Daniel Koll for sharing the code used to transform data to tidally locked coordinates. We are grateful to Tad Komacek, Dorian Abbot, and Peter Read for providing useful comments on an early version of this manuscript. M.H. was supported by a Lindemann Trust Fellowship from the English-Speaking Union. N.T.L was supported by Science and Technology Facilities Council grant no. ST/S505638/1.}

\showacknow{} % Display the acknowledgments section

% Bibliography
\bibliography{main}

\end{document}